\documentclass[lettersize,journal]{IEEEtran}
\usepackage{amsmath,amsfonts}
\usepackage{algorithmic}
\usepackage{algorithm}
\usepackage{array}
\usepackage[caption=false,font=normalsize,labelfont=sf,textfont=sf]{subfig}
\usepackage{textcomp}
\usepackage{stfloats}
\usepackage{url}
\usepackage{verbatim}
\usepackage{graphicx}
\usepackage{cite}
\usepackage{booktabs}
\usepackage{multirow}
\usepackage[table]{xcolor}
\usepackage{amssymb}
\def\BibTeX{{\rm B\kern-.05em{\sc i\kern-.025em b}\kern-.08em
    T\kern-.1667em\lower.7ex\hbox{E}\kern-.125emX}}
\usepackage{balance}

\usepackage{graphicx} 
\usepackage{hyperref} 
\usepackage[all]{hypcap}
\usepackage{cleveref}

\usepackage{xcolor}
\usepackage{soul}      
\soulregister\cite7
\soulregister\Cite7
\soulregister\parencite7
\soulregister\textcite7
\soulregister\cref7
\soulregister\Cref7
\soulregister\ref7
\soulregister\eqref7

\soulregister\cred7

\usepackage{ifthen}

\newif\ifrevmode
\revmodetrue          

\newcommand{\orcidicon}{\includegraphics[scale=0.6]{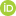}} 

\newcommand{\orcidlink}[1]{\href{https://orcid.org/#1}{\orcidicon}}

\begin{document}

\title{Universal Vessel Segmentation for Multi-Modality Retinal Images}
\author{Bo Wen \raisebox{1.5pt}{\orcidlink{0009-0001-5747-4739}}, \IEEEmembership{Graduate Student Member, IEEE}, Anna Heinke \raisebox{1.5pt}{\orcidlink{0000-0002-7115-8767}}, Akshay Agnihotri \raisebox{1.5pt}{\orcidlink{0009-0001-4624-8250}}, \\ Dirk-Uwe Bartsch \raisebox{1.5pt}{\orcidlink{0000-0003-0955-8708}}, William Freeman \raisebox{1.5pt}{\orcidlink{0000-0001-9979-2500}}, Truong Nguyen \raisebox{1.5pt}{\orcidlink{0000-0002-5022-063X}}, \IEEEmembership{Fellow, IEEE}, Cheolhong An \raisebox{1.5pt}{\orcidlink{0000-0003-2821-7386}} 
}

\definecolor{lightblue}{rgb}{0.8,0.9,1}
\definecolor{lightgreen}{rgb}{0.8,1,0.9}
\definecolor{lightred}{rgb}{0.9,0.6,0.6}



\maketitle

\crefname{table}{Table}{Tables}
\crefname{figure}{Fig.}{Figs.}

\begin{abstract}
We identify two major limitations in the existing studies on retinal vessel segmentation: (1) Most existing works are restricted to one modality, i.e., the Color Fundus (CF). However, multi-modality retinal images are used every day in the study of the retina and diagnosis of retinal diseases, and the study of vessel segmentation on other modalities is scarce; (2) Even though a few works extended their experiments to new modalities such as the Multi-Color Scanning Laser Ophthalmoscopy (MC), these works still require fine-tuning a separate model for the new modality. The fine-tuning will require extra training data, which is difficult to acquire. In this work, we present a novel universal vessel segmentation model (URVSM) for multi-modality retinal images. In addition to performing the study on a much wider range of image modalities, we also propose a universal model to segment the vessels in all these commonly used modalities. While being much more versatile compared with existing methods, our universal model also demonstrates comparable performance to the state-of-the-art fine-tuned methods. To the best of our knowledge, this is the first work that achieves modality-agnostic retinal vessel segmentation and the first to study retinal vessel segmentation in several novel modalities. \footnote{Code, model and 3 new retinal vessel segmentation datasets are available at \textbf{https://github.com/JRC-VPLab/URVSM}} \footnote{This paper is accepted by IEEE Transactions on Image Processing and will be published soon.}
\end{abstract}


\section{Background and Introduction}
\begin{figure*}[ht]
\centering
\includegraphics[width=1\textwidth]{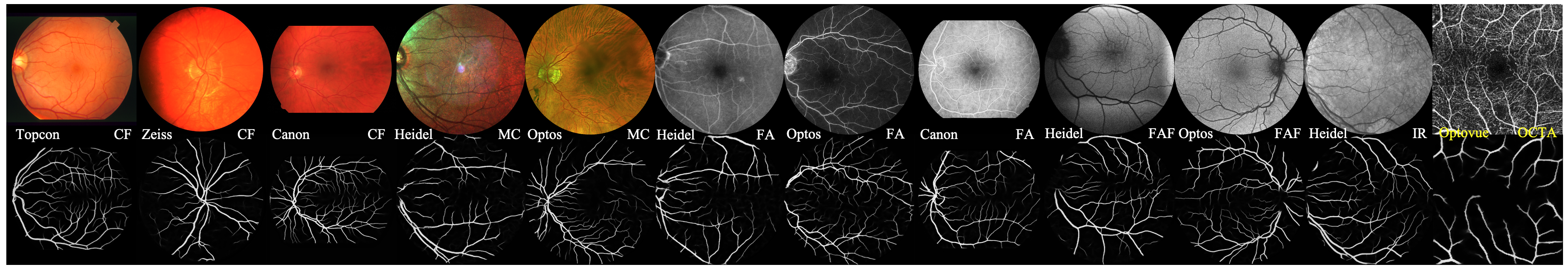}
\vspace{-0.7cm}
\caption{Overview of multi-modality retinal images and the segmentation results from our URVSM. In each image, the upper half is the original image and the lower half is the segmentation. In the middle, the camera (left) and modality (right) of the image is labeled. `Heidel' refers to Heidelberg Spectralis.}
\label{fig:intro}
\vspace{-0.5cm}
\end{figure*}

\IEEEPARstart{R}{etinal} vessel segmentation is a fundamental task in retinal image processing. It has numerous important applications, including retinal image registration \cite{Registration-Wang-TIP, Registration-Zhang-TIP}, human biometric identification \cite{Biometric-Alex}, retinal artery-vein classification \cite{AVCls-Estrada} and tree topology analysis in digital images \cite{TreeTopology-Estrada}. Clinically, the evaluation of retinal vessels is also important since the vessel features are predictive in several systemic diseases \cite{Clinical-Ikram, Clinical-Kawasaki, Clinical-McGee} and vessel segmentation is the key to the automation of the evaluation. However, most works in retinal vessel segmentation focus only on Color Fundus (CF) images. CF uses visible white light to illuminate the retina and the reflected light is used for generating the image. Nevertheless, although CF was the most widely used modality in the past, it requires dilation before imaging and is uncomfortable for patients. Moreover, the pathological information provided by CF is limited. Therefore, in more recent years, CF is used less frequently and emerging new imaging techniques become increasingly popular. These modalities include (\cref{fig:intro}): (1) Multi-Color Scanning Laser Ophthalmoscopy (MC), which uses a more intense laser beam and allows visualization of structures in deeper layers of the retina; (2) Fluorescence Angiography (FA), which uses fluorescent dye to illuminate the retinal vessels and is commonly used in the diagnosis of diseases related to neovascularization; (3) Fundus AutoFluorescence (FAF), which does not rely on an external light source but uses natural fluorescence of the retina and provides insights into the health and metabolism of the retinal pigment epithelium (RPE) and photoreceptor cells; (4) Infrared/Near-Infrared Reflectance (IR/NIR), which uses Infrared or Near-Infrared light to illuminate the image and allows visualization of deeper features on the retina, such as RPE abnormalities; (5) Optical Coherence Tomography Angiography (OCTA), which uses low-coherence interferometry to measure the backscattered light from retinal layers. OCTA has a much smaller imaging area than the other modalities and has a considerably different retinal vessel morphology and topology.

One major challenge that impedes the development of machine learning-based retinal vessel segmentation algorithms is the lack of data. Due to the complexity of the retinal vessels and the existence of many fine details, labeling is very time-consuming. Existing retinal vessel segmentation datasets (especially those with high label quality) are scarce and small, even for the CF modality, which usually only has 20-40 images. In addition, to the best of our knowledge, there are no publicly available vessel segmentation datasets for FAF and IR. For FA, there is only a very small public dataset with 8 images \cite{Dataset-Vampire}. In this work, we prepare three high-quality datasets for FA, FAF and IR, respectively. Each dataset has 40 images, and the vessels are carefully labeled by retinal experts. Nevertheless, despite the availability of the new datasets, in this work, we use them only for evaluation purposes. The reason is that we believe that there can never be enough annotated images to allow the training of a powerful universal model. Instead, we need to train the model with limited existing data and seek to overcome the challenge through innovations in the algorithm. In fact, our segmentation network is trained with only 19 annotated CF images from the widely used DRIVE \cite{Dataset_DRIVE} dataset. 

There are limited previous works that extend beyond the CF modality \cite{MMVesseg-Rodri, MMVesseg-Zhang, MMVesseg-StyleTransferPeng}. In addition to lacking comprehensive modalities (compared to our approach), one still needs to fine-tune a separate model on the target modality, where each model only works for one image modality. We instead aim to build a single universal model that achieves robust segmentation for all the commonly used modalities. The model could serve as an important tool in the cross-modality retinal image research, which becomes increasingly important in modern research of the retina. To achieve the universal segmentation with limited labeled data, we develop our model in two major steps (\cref{fig:pipeline}). First, we propose to use image translation as a preliminary domain adaptation method and train a model that translates input images from arbitrary modalities and cameras into a uniform Topcon (the Topcon TRC-50DX series is explicitly referred to as the gold standard in color fundus photography) CF image modality. We accessed an abundance of de-identified multi-modality retinal images from multiple widely used cameras (Topcon, Heidelberg Spectralis, Optos) at the Jacobs Retina Center at Shiley Eye Institute, University of California, San Diego (UCSD). These images allow us to train a useful image translation model in a self-supervised manner. Furthermore, we adopt data augmentation to account for the slight domain difference with the other CF cameras which our center does not have, including Canon and Zeiss. Secondly, we train a segmentation model on the Topcon CF domain to perform the segmentation. However, image translation cannot fully eliminate the domain gaps between different modalities given such a difficult many-to-one translation task. To solve this problem, we propose to use a topology-aware segmentation method that learns topological features of the vessels in addition to their conventional pixel-level features. The topology-aware segmentation methods were initially proposed to improve the topological accuracy of the segmentation of curvilinear structures. We go beyond their conventional applications and propose topological learning as a novel domain adaptation method for retinal vessel segmentation, in a way that forces the segmentation to be topologically correct, so when images from different domains are input, the model can still yield robust segmentation.

Our universal vessel segmentation model is evaluated on 10 diverse datasets (including 3 new datasets we prepared for this work) of the 6 most commonly-used retinal image modalities (CF, FA, FAF, MC, IR/NIR, OCTA) taken from 6 most commonly used camera systems (Topcon, Canon, Zeiss, Heidelberg Spectralis, Optos, Optovue). Our universal model is compared with state-of-the-art methods that are fine-tuned on each of the 7 datasets, respectively, and is shown to achieve comparable performance. Extensive ablation studies were also performed to investigate how different components in the proposed pipeline affect the performance of the universal vessel segmentation. The main contributions of this paper are:
\begin{itemize}
    \item We propose the universal retinal vessel segmentation task. To the best of our knowledge, this is the first work that treats retinal vessel segmentation as a modality-agnostic segmentation task covering multiple imaging modalities and camera systems, reflecting the real-world need for generalizable retinal vessel segmentation solutions.
    \item We propose the first universal (retinal) vessel segmentation model (URVSM) for multi-modality retinal images.
    \item We investigate image-to-image translation between retinal images from different modalities. We explore different Generative Adversarial Network (GAN)-based and Diffusion Model-based image translation methods and evaluate their performance on retinal images.
    \item We propose topological feature learning as a novel domain adaptation strategy for multi-modal retinal vessel segmentation. We show that by learning a combination of established topological features of the vessels, the segmentation model becomes much more robust to inputs from different domains (modalities and imaging systems).
    \item We release three new, high-quality vessel segmentation datasets for the FA, FAF, and IR retinal images, respectively.
    \item We release refined vessel annotations of the widely used DRIVE dataset with improved topological accuracy (to be discussed in \cref{sssec:segtrain_dataset}). 
\end{itemize}

\section{Related Work}

\subsection{Retinal Vessel Segmentation}
Retinal vessel segmentation is a specialized task that general-purpose foundational segmentation models \cite{SAM2, GroundedSAM2, MedSAM2} struggle to handle effectively. Current works in retinal vessel segmentation focus on segmenting vessels in Color Fundus retinal images, where various machine learning-based or non-machine learning-based methods were proposed. A majority of the early methods are based on match filtering \cite{Vesseg-MF-Chaudhuri, Vesseg-MF-Kovacs, Dataset-HRF}, which convolves retinal images with predefined or learned kernels and then adaptively thresholds the image to obtain the vessel segmentation map. Another type of methods use tracking algorithms to mathematically estimate the growth of the retinal vessel trees from seed points \cite{Vesseg-Tracking-Delibasis, Vesseg-Tracking-Lin}. The segmentation can then be derived by extending around the estimated vessel centerlines using the estimated vessel diameters. Some methods also adopt morphological image processing methods such as the top-hat operation to extract the vessels or as a post-processing step \cite{Vesseg-Morph-Graz, Vesseg-Morph-Imani}. Furthermore, numerous traditional machine learning-based methods were also applied in retinal vessel segmentation such as the support vector machine \cite{Vesseg-SVM-Tang}, random forest \cite{Vesseg-RF-Wang} and Adaboost \cite{Vesseg-Adaboost-Memari} for supervised methods, and the Gaussian mixture model \cite{Vesseg-GMM-Roy}, Fuzzy C-Means \cite{Vesseg-FCM-Neto} for unsupervised methods. In more recent years, deep learning-based methods dominate the literature. They mainly rely on the Convolutional Neural Networks (CNN) as a backbone models due to the existence of vast fine details and the spread-out characteristics of the vessels. An introductory work is the Deep Retinal Image Understanding (DRIU) \cite{Vesseg-CNN-DRIU}, which adopts a VGG \cite{Others-VGG} network structure as an encoder and assigns specialized output layers to the embedded features to segment the vessels and the optic disc from the retina. Based on a widely used U-Net \cite{Others-U-Net} network structure and to seek further improvement for retinal vessel segmentation, \cite{Vesseg-CNN-CENet, Vesseg-CNN-CS2Net, Vesseg-CNN-GlobalTransformer, Vesseg-CNN-MCDAUNet, Vesseg-CNN-MCGNet} proposed to add various global reasoning modules at the bottom of the network. \cite{Vesseg-CNN-MCDAUNet, Vesseg-CNN-MCGNet, Vesseg-CNN-RCARUNet, Vesseg-CNN-WaveNet, Vesseg-CNN-GlobalTransformer} proposed to add feature aggregation modules in the residual connection at each level of the network. \cite{Vesseg-CNN-DUNet, Topo-TCLoss} proposed to use deformable convolution where the convolution kernel shapes are learned to adapt to the elongated morphology of the retinal vessels. With similar motivation, \cite{Vesseg-CNN-WSDMF} proposed an orientation-selective convolution to learn selective receptive fields of the kernel to adapt to the shape of the vessels. \cite{Vesseg-CNN-GlobalTransformer} proposed to use improved feature fusion methods to better preserve the fine details of vessels in the segmentation. In addition to a single U-Net-like segmentation network, several works \cite{Vesseg-CNN-WNet, Vesseg-CNN-WSDMF, Vesseg-CNN-FANet} proposed to use cascaded networks or a single looped network to iteratively refine the segmentation. 

\subsection{Unpaired Image-to-Image Translation} 
Image-to-image translation aims to transfer a source image to a target image domain while preserving its structural information. However, it is difficult to prepare paired retinal images from the source and target domains for training. Therefore, in this work, we look into the unpaired image-to-image translation methods where the projection between the two domains can be learned with unpaired images with no structural correspondences. A representative and most commonly used method in this task is CycleGAN \cite{Trans-CycleGAN}, which introduces a cycle consistency that projects the translated image back to the source domain to enforce retaining the structural information in the translated image. Following works further introduce geometric consistency \cite{Trans-CcGAN}, mutual information regularization \cite{Trans-Mutual}, and contrastive unpaired translation \cite{Trans-CUT} to improve the computational efficiency over the cycle consistency-based methods during the training phase. \cite{Trans-MUIST, Trans-StarGAN} proposed one-to-many transfer schemes which allow the translation of images to multiple domains. While the above works are based on GAN backbones and achieve good image translation performance, diffusion models \cite{Others-Diffusion, Others-DDPM, Others-DDIM} have recently gained more research attention on this topic for easier training and superior image quality. However, the methods \cite{Trans-DDIB, Trans-ENOT, Trans-SynDiff} are computationally intensive and were only applied to small images. For example, \cite{Trans-SynDiff} combines a diffusion model with adversarial learning and formulates the training in a cycle-consistency framework, which requires training two diffusion models at the same time. Most recently, \cite{Trans-UNSB} proposed an Unpaired Neuron Schrödinger Bridge (UNSB) with a novel scheme to simulate the Schrodinger Bridge, which is a random process that interpolates between two image domains via a diffusion process. It has higher computational efficiency and makes the application to higher-resolution images (e.g., retinal images) possible.

Image style transfer \cite{Trans-NAAS, Trans-PerceptualLoss, Trans-EFDM, Trans-StyTr2, Trans-StyleDiffusion} is another method which can be seen as a special type of unpaired image-to-image translation but with a perceptual inductive bias. The majority of the methods use a perceptual feature space to compute a style loss and inject the style of a reference image into the input content image, while using another self-comparison content loss to retain the content information of the input in the transferred image. We also explore methods in this topic since style transfer is used for non-CF retinal vessel segmentation in \cite{MMVesseg-Zhang, MMVesseg-StyleTransferPeng}.

\subsection{Topology-Aware Image Segmentation}
Topology-aware image segmentation aims to preserve the topological accuracy in the segmentation of tubular and net-like structures. The literature mainly considers a topological loss function that extracts topological features from the ground truth and segmentation and minimizes the difference between them. One type of methods extract certain features from the segmentation which approximate the actual topological features of the foreground objects. A pioneering work \cite{Topo-PTLoss} uses an ImageNet-pretrained VGG \cite{Others-ImageNet, Others-VGG} network after the segmentation network. The perceptual features extracted from different layers of the pretrained VGG network are shown to be approximations of the topological features. The topological loss is then computed between the ground truth and predicted perceptual features. Another topological segmentation loss function \cite{Topo-GlandSeg} was proposed exclusively for the segmentation of glands. The loss is computed according to the iterations needed to derive the gland skeletons by the image erosion operation. \cite{Topo-clDice} proposed to approximate the topological features of tubular structures by a relationship between the foregrounds in the ground truth and prediction and their soft-skeletons. Using such a relationship as a loss function is shown to strengthen the connectivity along the tubular structure, thus providing better topological accuracy. 

Another type of methods use strict topological features of the segmentation which are based on the computational topology theory and have received more research attention in recent years. The topological features are mainly based on the persistent homology theory \cite{Topo-WTLoss, Topo-WTLoss3D, Topo-CMR,  Topo-TPAMI, Topo-TCLoss, Topo-MLLoss, Topo-BMLoss, Topo-SATLoss}. A revolutionary work that first introduced this concept in image segmentation is \cite{Topo-WTLoss}, which computes the persistent features from the ground truth and prediction, matching the features using a Wasserstein distance \cite{Topo-Wassersteinforpersistence} and using the distance as the loss function. The following works share a similar idea. \cite{Topo-WTLoss3D} also uses the Wasserstein distance but applies the method on 3D point cloud data. \cite{Topo-MLLoss} proposed to match the longest persistent barcodes in the prediction to the ground truth. \cite{Topo-TCLoss} proposed to use a Hausdorff distance \cite{Topo-Hausdorff_distance} to match the persistent diagrams to improve the sensitivity of the matching to the outliers. \cite{Topo-SATLoss} and \cite{Topo-BMLoss} both aim to compute more precise persistent feature matching, as in earlier methods, the actual matching results are highly inaccurate due to only considering the global topological features. \cite{Topo-SATLoss} and \cite{Topo-BMLoss} instead allow the loss to act more selectively on the topological features thus improves the efficiency of the topological loss function. \cite{Topo-BMLoss} proposed to use an Induced Matching \cite{Topo-Induced_matching} method which constructs a common ambient space between the ground truth and prediction and maps the persistent features in between to find a better matching. However, the computational cost of this method is very high, especially for large images, where typical retinal images usually have high resolutions. Alternatively, \cite{Topo-SATLoss} proposes to leverage the spatial locations of the persistent features referencing their critical cell locations for more faithful persistent feature matching, and is computationally much more efficient than \cite{Topo-BMLoss}.

\begin{figure*}[ht]
\centering
\includegraphics[width=1\textwidth]{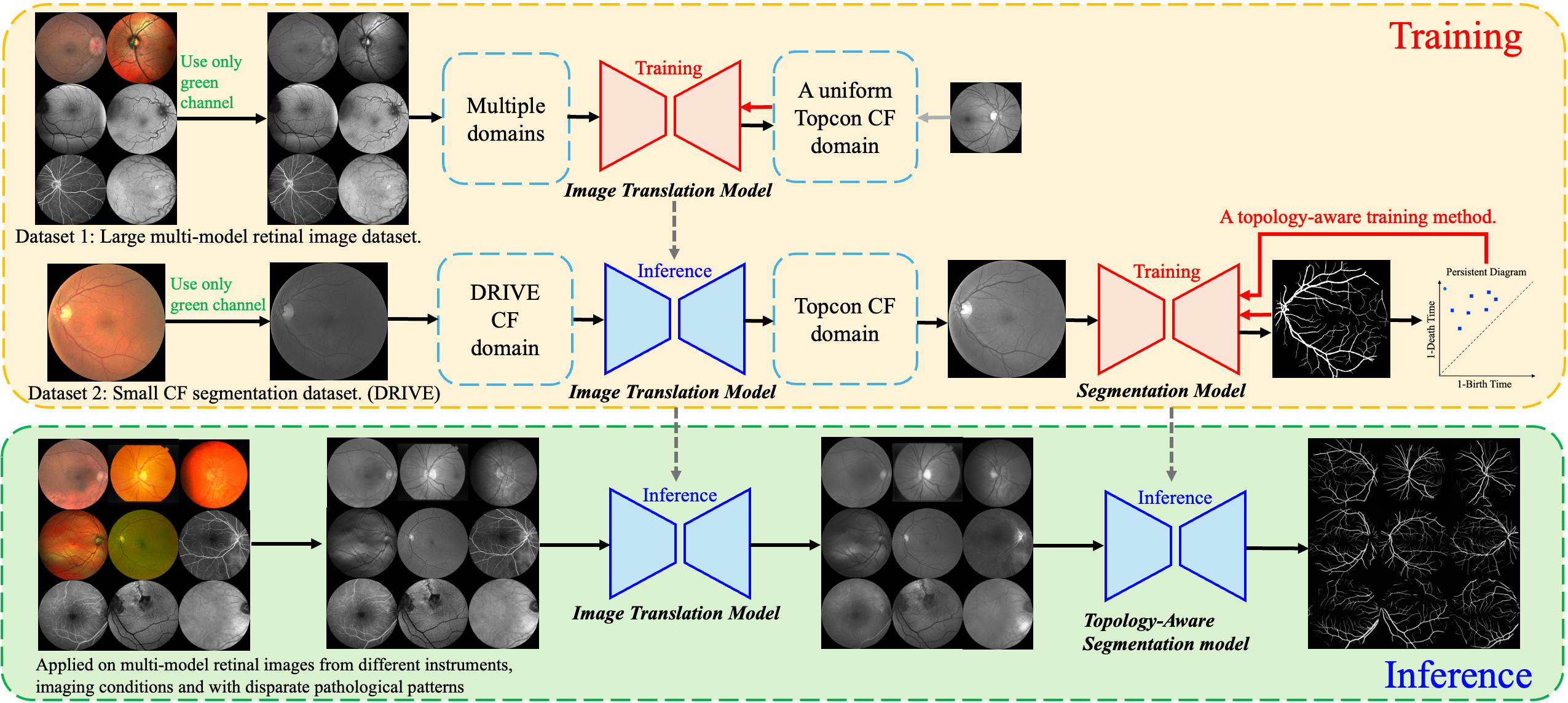}
\vspace{-0.7cm}
\caption{Pipeline for our proposed universal retinal vessel segmentation model, which includes the 2-stage training and the inference.}
\label{fig:pipeline}
\vspace{-0.6cm}
\end{figure*}

\section{Universal Retinal Vessel Segmentation Model}
\label{sec:overview}

\subsection{Overview}
We discuss the overall training and inference pipeline (\cref{fig:pipeline}) in this subsection and provide more details about the major components in the pipeline in the following sub-sections. Given multi-modality retinal images, we first adopt a simple yet widely used preprocessing method that takes only the green channel from the RGB image. All of the training and inference are then performed on the green-channel images. 

Following preprocessing, we train a domain adaptation model for multi-modality retinal images. Although several domain adaptation methods were proposed for retinal vessel segmentation, such as using pseudo-label methods \cite{Vesseg-CNN-WNet, MLAL-PDSF},
teacher-student networks \cite{FRR-TSNet}, adversarial learning \cite{Vesseg-CNN-MCGNet, MLAL-PDSF} and asymmetrical maximum classifier discrepancy \cite{UA-Vesseg-AMCD}, they were proposed only to address the domain gap between different datasets (from different cameras) in CF images. Moreover, these methods still require unsupervised fine-tuning on the target domain images, which deviate from our goal of universal segmentation. Therefore, in this work we propose to use image translation as the preliminary domain adaptation method, which allows us to train a single model for all the modalities.

The image translation model aims to translate images from all the modalities to a uniform Topcon CF modality, and is trained between an equivalent number of non-CF images (FA, FAF, MC, IR/NIR) and the Topcon CF images, i.e., during the training, we treat all the non-CF modalities as a single image domain. More information on the training dataset is provided in \cref{sssec:trans_dataset}. Furthermore, the translation network is trained with an identity loss (will be discussed in detail in \cref{subsec:image_translation}), which allows us to use a single translation network to transfer all the images to the Topcon CF domain without the need to specify whether they are CF or non-CF. In addition, the translation model also translates CF images from other cameras to Topcon.

After the image translation model is trained, we use it to translate our segmentation training set (19 images from the DRIVE \cite{Dataset_DRIVE} training set with vessel labels) to the Topcon CF domain. The translated images are combined with the original images as an augmented training set to train the downstream segmentation network. Nevertheless, since we are training a difficult many-to-one (FA, FAF, MC, IR/NIR to CF) translation, the translation model cannot perfectly address the domain gap between the non-CF images and the CF images (\cref{fig:ab_transfer_examples}). To solve this problem, we propose topology-aware segmentation as a novel domain adaptation method. We show that by learning additional topological features of the vessels, the segmentation model becomes much more robust to inputs from different domains. In practice, in addition to a conventional pixelwise loss function (BCELoss), the segmentation network is jointly trained with two topological loss functions.

At inference, we can disregard the modality of the image. Given a retinal image from an arbitrary modality, we simply need to (1) extract the green channel; (2) apply the translation network to the green channel image; (3) apply the topology-aware segmentation network to the translated image to obtain the final segmentation.

\subsection{Image Translation Network}
\label{subsec:image_translation}

Multiple GAN-based and diffusion model-based image translation methods are explored for our task. For the diffusion-based method, we experiment with StyleDiffusion \cite{Trans-StyleDiffusion} and Unpaired Neuron Schrödinger Bridge (UNSB) \cite{Trans-UNSB}. For the GAN-based method, we experiment with CycleGAN \cite{Trans-CycleGAN}. We will show in \cref{sssec:ab_image_translation} that CycleGAN, despite being a classic method, outperforms the other two most recent methods in most datasets. Therefore, we pick CycleGAN as the backbone of our image translation model. Consequently, the following sections discuss the implementation based on CycleGAN, and we refer the readers to \cite{Trans-StyleDiffusion, Trans-UNSB} for details about the other two methods and to \cref{sssec:ab_image_translation} for our implementation details.

\begin{figure}[]
\centering
\includegraphics[width=0.45\textwidth]{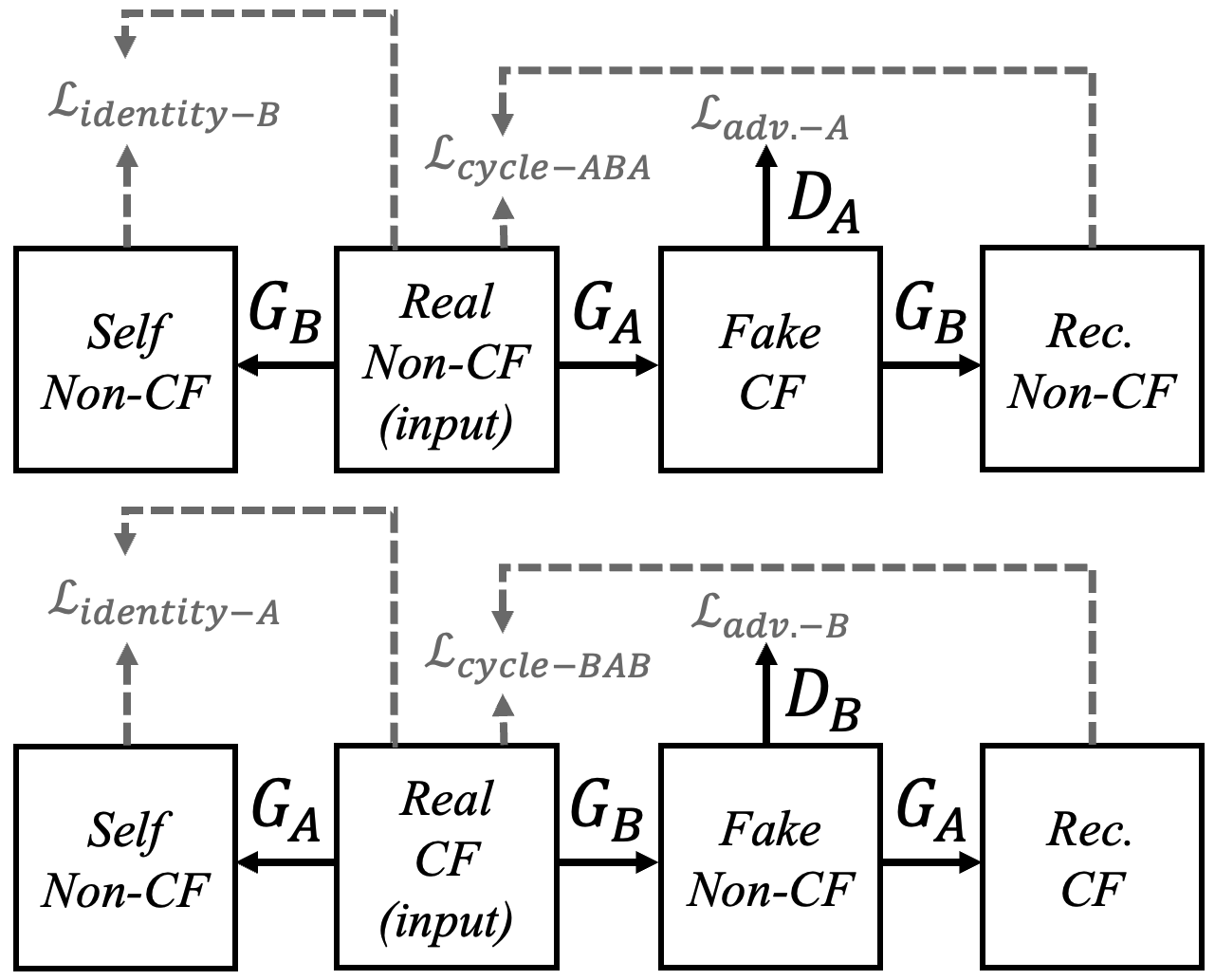}
\vspace{-0.3cm}
\caption{Overview of our image translation network, which is based on CycleGAN \cite{Trans-CycleGAN}. $G$ refers to the GAN generators and $D$ for the discriminators. The gray dotted lines illustrate how loss functions are computed between different images.}
\label{fig:CycleGAN}
\vspace{-0.6cm}
\end{figure}

To train the translation network, we treat all the non-CF images (FA, FAF, MC, IR/NIR) as image type A and the Topcon CF images as image type B. At each iteration, one batch of image type A and one batch of image type B are selected from the training set. As illustrated by \cref{fig:CycleGAN}, generator A is trained to translate all the type A images (non-CF) to a fake type B (CF), and generator B is used to translate type B to a fake type A, which is a neutral non-CF domain. Both generators A and B are learned with adversarial learning \cite{Others-GAN, Others-LSGAN}, where the generator and discriminator are alternately updated. At each iteration, given the image batch $I_{A}$, the generators (A+B) are first updated with the objectives (discriminators frozen):
\begin{equation}
    \textit{L}_{G_{A}} = \textit{L}_{adv.-G_{A}} + \lambda_{c}\textit{L}_{cycle-A} + \lambda_{i}\textit{L}_{identity-A}
\end{equation}
where the adversarial loss is defined as:
\begin{equation}
    \textit{L}_{adv.-G_{A}} =  \left | D_{A}(G_{A}(I_{A})) - 1 \right |^{2}_{2}
\end{equation}
Furthermore, the translated fake CF images are again fed into the opposite generator to reconstruct the non-CF images. The reconstruction is supervised by a cycle consistency loss:
\begin{equation}
    \textit{L}_{cycle-A} =  \left | I_{A} - G_{B}(G_{A}(I_{A})) \right |_{2}
\end{equation}
and is controlled by a weight $\lambda_{c}$. Finally, the real images are also passed through their opposite generator to generate the image itself. This step is supervised with an identity loss:
\begin{equation}
    \textit{L}_{identity-A} =  \left | I_{A} - G_{B}(I_{A}) \right |_{2}
\end{equation}
and is controlled by a weight $\lambda_{i}$. The identity loss was initially proposed to correct the tint in the translated image \cite{Trans-CycleGAN}. However, we find it especially useful in our task since it allows us to use the generator $G_{A}$ to map the CF images to itself. As a result, at inference we only need to apply one generator $G_{A}$ to all the input images regardless of their modalities. If the image is non-CF, the translation model will convert it to CF. If the image is already CF, the output image will remain a CF image. It allows our universal model to work without the need to provide any modality information about the input.

After the generator is updated, the discriminator is updated with the adversarial loss (generators frozen):
\begin{equation}
    \textit{L}_{adv.-D_{A}} =  \frac{1}{2}\left | D_{A}(I_{B}) - 1 \right |^{2}_{2} + \frac{1}{2}\left | D_{A}(G_{A}(I_{A})) \right |^{2}_{2}
\end{equation}

For the opposite image batch $I_{B}$, the generators (B+A) and discriminator (B) are updated in the same way, as illustrated in \cref{fig:CycleGAN}. In addition, we first update the generators for both $I_{A}$ and $I_{B}$, and then update the discriminators.

\subsection{Topology-aware Segmentation Network}
\label{subsec:topological_segmentation}

The downstream segmentation network is learned with a novel combination of the Perceptual Topological Loss $\mathcal{L}_{tc}$ and the Persistent Topological Loss $\mathcal{L}_{ts}$, introduced below.

\subsubsection{Perceptual Topological Loss}
\label{sssec:perceptual_topological_loss}
\begin{figure}[ht]
\centering
\includegraphics[width=0.4\textwidth]{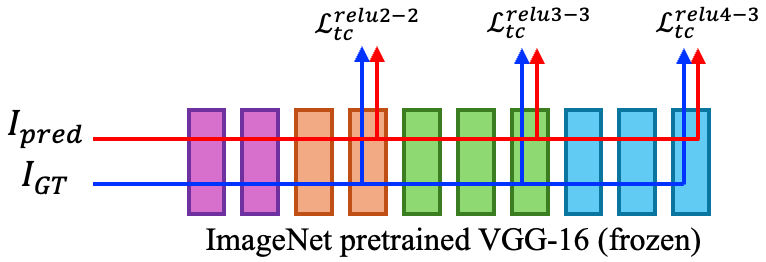}
\vspace{-0.3cm}
\caption{Illustration of the perceptual topological loss function.}
\label{fig:topo_perceptual}
\vspace{-0.1cm}
\end{figure}

We first propose to use a perceptual topological loss \cite{Topo-PTLoss} where the topological information is approximated by the perceptual features encoded by an ImageNet \cite{Others-ImageNet}-pretrained VGG-16 \cite{Others-VGG} network. As illustrated by \cref{fig:topo_perceptual}, image embeddings $\mathcal{E}_{l}(\cdot)$ are computed from the predicted vessel probability map $I_{pred}$ and the ground truth $I_{GT}$, where $l$ denotes different layers in the VGG-16 network. In practice, we take the output of the $relu2\_2$, $relu3\_3$, and $relu4\_3$ layers. Then, the perceptual topological loss function $\mathcal{L}_{tc}$ is defined as:
\begin{equation}
    \mathcal{L}_{tc} = \sum_{l}^{N_{l}}\left | \mathcal{E}_{l}(I_{pred}) - \mathcal{E}_{l}(I_{GT}) \right |^{2}_{2}
\end{equation}
\vspace{-0.3cm}

\subsubsection{Persistent Topological Loss}
\label{ssec:persistent_topological_loss}

\begin{figure}[]
\centering
\includegraphics[width=0.47\textwidth]{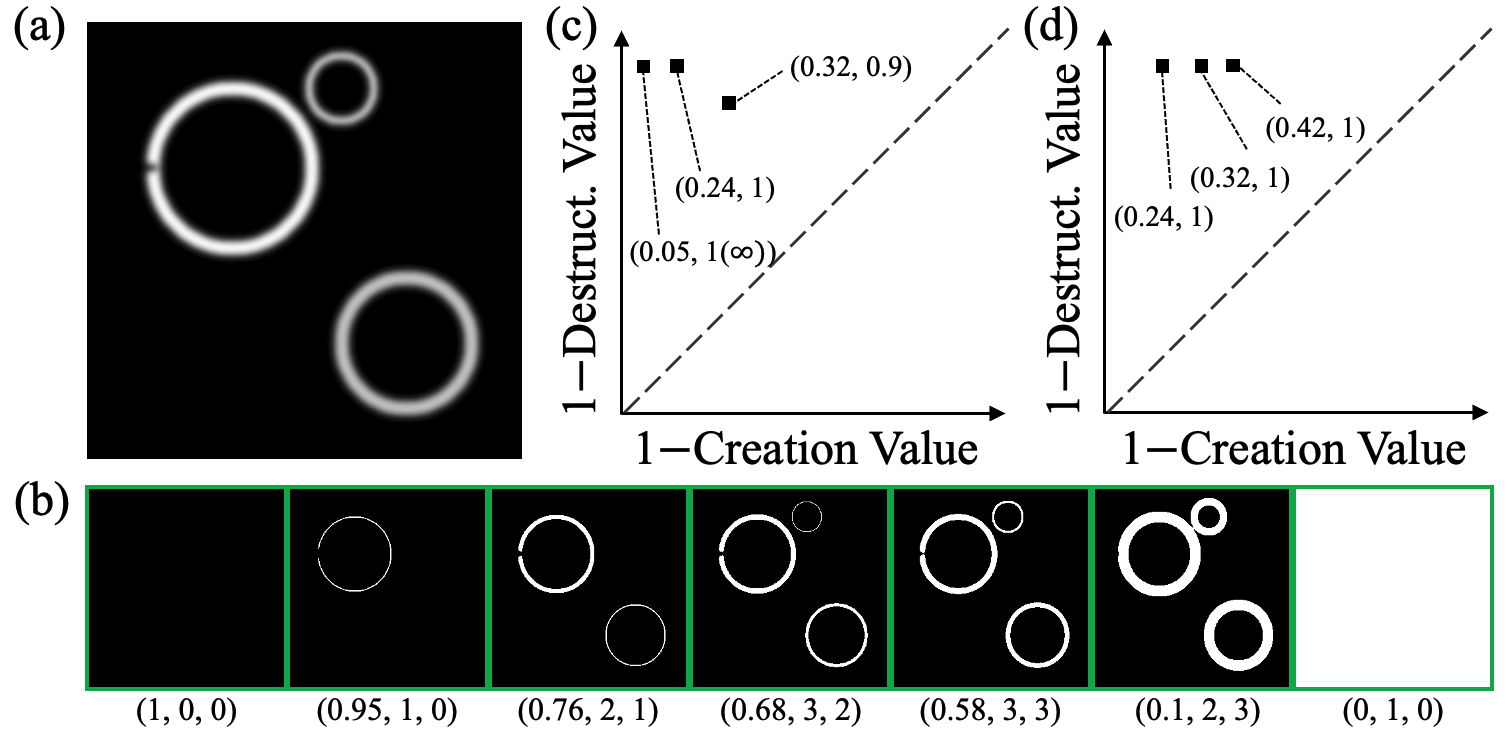}
\vspace{-0.3cm}
\caption{A toy example visualizing persistent homology in images: (a) a grayscale image; (b) the filtration visualized by thresholded binary images. Under each image, the digits represent the filtration value (threshold value), the number of 0-homology features, and the number of 1-homology features; (c) the 0-persistent diagram; (d) the 1-persistent diagram.}
\label{fig:PH}
\vspace{-0.6cm}
\end{figure}

In a 2D digital image, the topology is described by its 0- and 1-homology features, i.e., the connected components and the loops. The persistent homology is a method to describe a process where homology features are created and destroyed in a sequence (called `filtration') of a cubical complex \cite{Topo-cc} (which encodes an image and the higher-dimensional relationships between its pixels as structured cells) and its subcomplexes determined by the values of cells. Although it deviates from the formal definition and how persistent homology is computed, a simpler way to interpret persistent homology in 2D digital images is by thresholding the image from its highest pixel value to its lowest pixel value and stopping at every pixel value that appears in the image. Each thresholded image is a binary image, where we can determine the connected components and the loops inside. As demonstrated by the example in \cref{fig:PH}, when the threshold value decreases, new pixels are added to the binary image, leading to the creation of new homology features or destruction of old features. A homology feature that `lives' through the decreasing thresholds is then called a `persistent feature'. We refer readers to \cite{Topo-CompTopoBook} for more details and strict definitions.

The persistent homology of an image can be described by a persistent diagram, where each point on the diagram represents one persistent feature, whose coordinates are (1$-$Creation Value, 1$-$Destruction Value). 

We propose to use a Spatial-Aware Topological Loss Function (SATLoss) \cite{Topo-SATLoss} as the persistent topological loss function. After computing persistent diagrams $\mathcal{D}(I_{pred})$ and $\mathcal{D}(I_{GT})$, the persistent features are matched using a spatially-weighted Wasserstein distance \cite{Topo-Wassersteinforpersistence}:

\begin{align}
     & W_{q}(\mathcal{D}(I_{pred}),\mathcal{D}(I_{GT}))= \\
     & [\inf_{\eta :\mathcal{D}(I_{pred})\rightarrow \mathcal{D}(I_{GT})}\sum_{p\in \mathcal{D}(I_{pred})}^{} \notag s_{p} \left \|p - \eta(p))  \right \|^{q}_{2}]^{\frac{1}{q}}
\label{eq:matching}
\end{align}
where $s_{p}=\left \|c_{b}(p)-c_{b}(\eta(p))  \right \|^{q}$ is a spatial weight that is computed between the creation cells ($c_{b}(\cdot)$) of the persistent features. $p$ and $\eta(p)$ denote a persistent feature from $I_{pred}$ and a candidate persistent feature from $I_{GT}$ to be matched. $q$ is the norm and in practice we take $q=2$. After finding the optimal matching by solving the optimal transport problem, the persistent topological loss function is computed as (reformulating $p$ and $\eta(p)$ by their creation and destruction values):
\begin{equation}
  \mathcal{L}_{ts} = \sum_{p \in \mathcal{D}(I_{pred})}^{}s^{*}_{p}([b(p) - b(\eta^{*}(p))]^{2}+[d(p) - d(\eta^{*}(p))]^{2})
\label{eq:satloss}
\end{equation}
where $\eta^{*}(p)$ is the optimal matching of $p$ in $\mathcal{D}(I_{GT})$ and $s^{*}_{p}$ is the corresponding spatial weight. $b(\cdot)$ and $d(\cdot)$ are the creation and destruction values of the topological features. The values are taken from $I_{pred}, I_{GT}$ instead of $\mathcal{D}(I_{pred}), \mathcal{D}(I_{GT})$ to make the loss differentiable \cite{Topo-SATLoss}. 

Furthermore, for $p \in \mathcal{D}(I_{pred})$ which are not matched with a point in $\mathcal{D}(I_{GT})$, we match it to the closest point on the diagonal of the persistent diagram, i.e., $b(\eta^{*}(p))$ and $d(\eta^{*}(p))$ equals to 1, 0 if the matched $\eta^{*}(p)$ is a point in $\mathcal{D}(I_{GT})$, otherwise:
\begin{equation}
    b(\eta^{*}(p)) = d(\eta^{*}(p)) = \frac{1}{2}(b(p)+d(p))
\end{equation}
The aforementioned matching process is based on all the candidate matchings, including matching with points in $\mathcal{D}(I_{GT})$ and matching with the diagonal.

\begin{figure}[ht]
\centering
\includegraphics[width=0.48\textwidth]{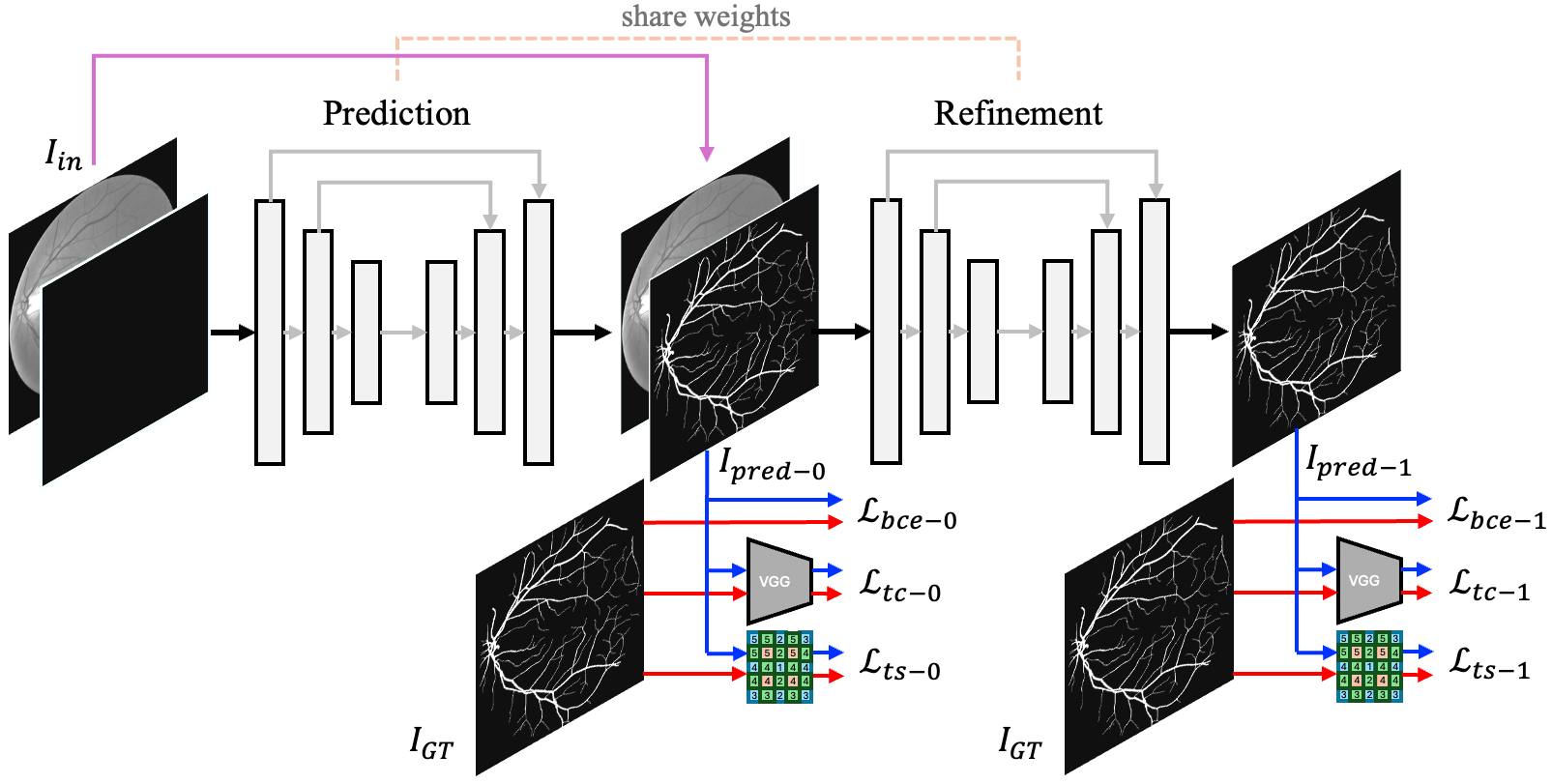}
\vspace{-0.3cm}
\caption{Overview of the segmentation network.}
\label{fig:segnet}
\vspace{-0.4cm}
\end{figure}

\subsubsection{Loss Formulation}
\label{sssec:loss_formulation}
Finally, the persistent topological loss and the perceptual topological loss are used with a conventional pixelwise BCELoss:

\begin{equation}
\mathcal{L}_{bce} = -\frac{1}{n}\sum_{i=1}^{n}y_{i}log(x_{i})+(1-y_{i})log(1-x_{i})
\end{equation}
where $x_{i}$ and $y_{i}$ are the per-pixel predicted likelihood values and the ground truth value. $n$ is the total number of pixels in the image. The topological loss functions are controlled by weights $\lambda_{tc}$ and $\lambda_{ts}$:

\begin{equation}
\mathcal{L}_{total} = \mathcal{L}_{bce} + \lambda_{tc}\mathcal{L}_{tc} + \lambda_{ts}\mathcal{L}_{ts}
\end{equation}
to jointly optimize the segmentation network.

\begin{figure}[]
\centering
\includegraphics[width=0.48\textwidth]{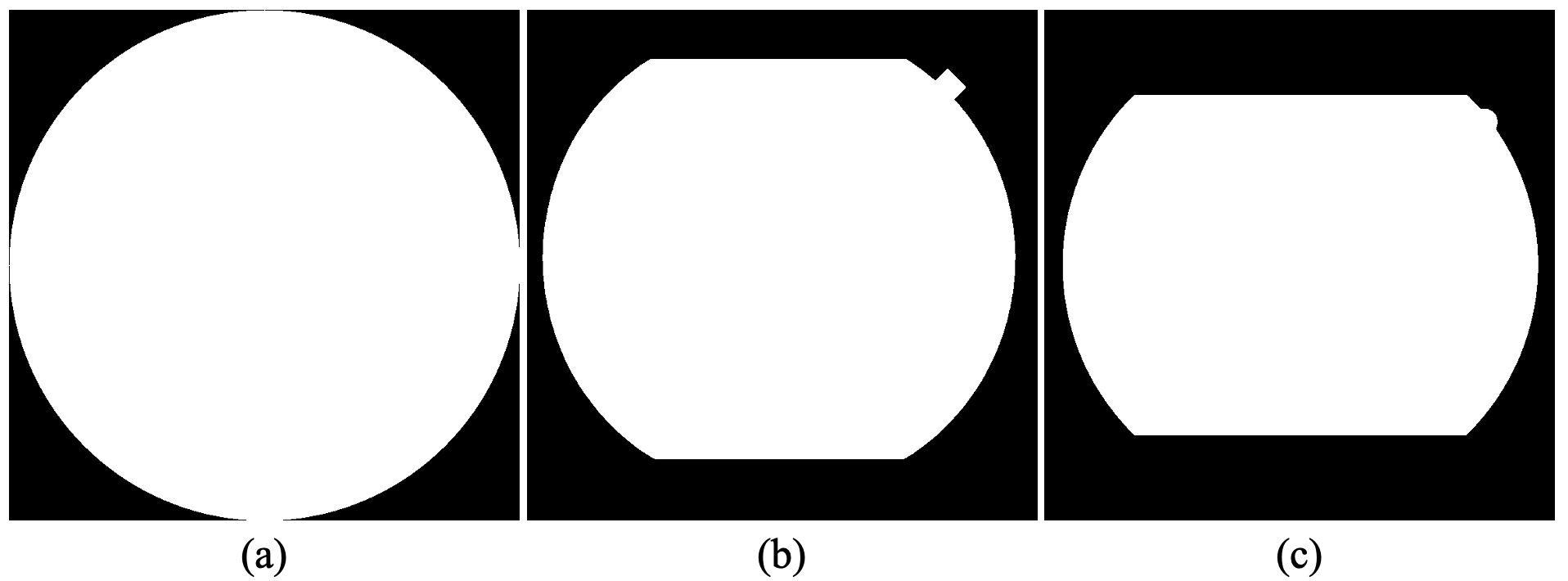}
\caption{Different types of masks randomly applied to the retinal images for training both the translation and segmentation networks.}
\label{fig:masks}
\end{figure}

\subsubsection{Segmentation Network}
\label{sssec:segnet}
We use a basic U-Net \cite{Others-U-Net} structure with reduced model size for the segmentation network. In addition, as shown in \cref{fig:segnet}, we adopt a `predict-refine' strategy, which first makes a coarse prediction of the vessels and feeds the initial prediction and the input image to the network again to refine it. This method has been shown to be useful in both topology-aware image segmentation \cite{Topo-PTLoss} and multiple works in deep learning-based retinal vessel segmentation \cite{Vesseg-CNN-WSDMF, Vesseg-CNN-WNet, Vesseg-CNN-FANet}. In our method, the prediction and refinement networks share the same weights. Additionally, at both the initial prediction stage and the refinement stage, the loss functions are computed to optimize the network.

\section{Experiments}
\label{sec:experiments}
\begin{table}
  \caption{Details of evaluation datasets}
  \vspace{-0.1cm}
  \label{tab:datasets_details}
  \centering
  \scriptsize

  \begin{tabular}{@{\hspace{2pt}}c@{\hspace{2pt}}|ccc@{}}
  \toprule
     Dataset & modality & camera & No. images \\
  \midrule
  \multicolumn{1}{c|}{STARE} & CF & Topcon & 20  \\
  \multicolumn{1}{c|}{HRF} & CF & Canon & 35  \\
  \multicolumn{1}{c|}{ChasedDB1} & CF & Zeiss & 28 \\
  \multicolumn{1}{c|}{IOSTAR} & MC & Heidelberg & 30 \\
  \multicolumn{1}{c|}{JRCFA} & FA & Heidelberg \& Optos & 40 \\
  \multicolumn{1}{c|}{JRCFAF} & FAF & Heidelberg \& Optos & 40 \\
  \multicolumn{1}{c|}{JRCIR} & IR & Heidelberg & 40 \\
  \multicolumn{1}{c|}{OCTA500-3M} & OCTA & Optovue & 40 \\
  \multicolumn{1}{c|}{OCTA500-6M} & OCTA & Optovue & 100 \\
  \multicolumn{1}{c|}{PRIME-FP20} & MC(UWF) & Optos & 15 \\

  \bottomrule
  \end{tabular}
  \vspace{-0.4cm}
\end{table}

\begin{table}
  \caption{Training (one-time) and inference (per item, image size 768 $\times$ 768) time of the URVSM: translation (Trans.) + segmentation (Seg.). Device: i9-13900K CPU + 1 RTX 4090 GPU.}
  \label{tab:comp_cost}
  \centering
  \scriptsize

  \begin{tabular}{cc|cc}
  \toprule
    Trans. Train & Seg. Train & Trans. Inference & Seg. Inference \\
  \midrule
   6.9h & 11.0h & 0.15s & 0.07s \\

  \bottomrule
  \end{tabular}
  \vspace{-0.6cm}
\end{table}

\begin{figure*}[htbp]
\centering
\includegraphics[width=0.99\textwidth]{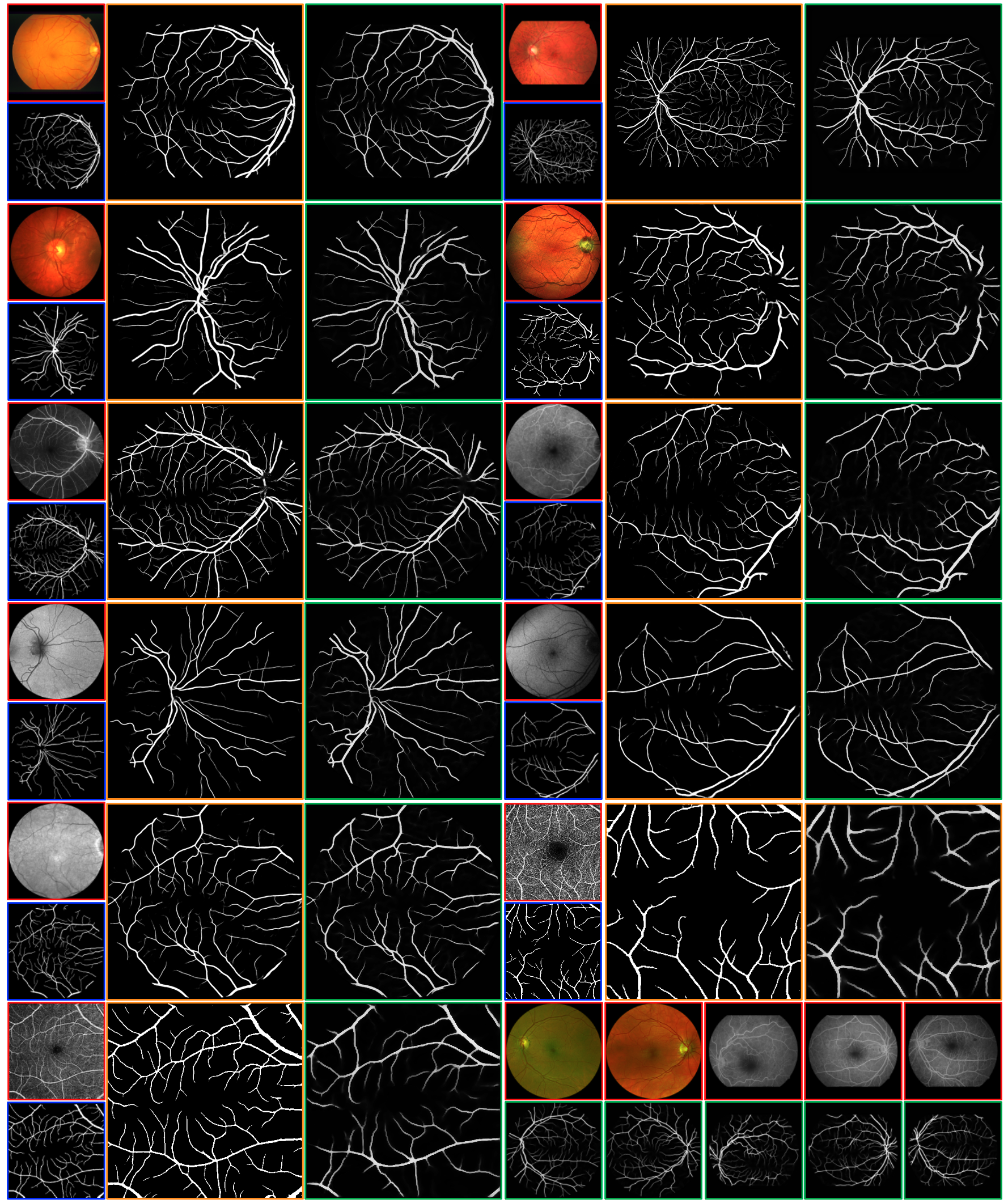}
\vspace{-0.3cm}
\caption{Qualitative comparison between our URVSM and fine-tuned SOTA models on multi-modality retinal images (best view with zooming in). Red, blue, orange and green box include the original image, the ground truth label, fine-tuned FA-Net segmentation and the segmentation from our URVSM. From left to right, top to down: Topcon CF, Canon CF, Zeiss CF, Heidelberg MC, Optos FA, Heidelberg FA, Optos FAF, Heidelberg FAF, Heidelberg IR. Optovue OCTA 3mm, Optovue OCTA 6mm, Bottom right shows the qualitative results for Optos MC and Canon FA (no ground truth available to compare with and to train the baseline fine-tuned models).}
\label{fig:main_quali}
\vspace{-0.6cm}
\end{figure*}

\begin{table*}[ht]
  \caption{Comparison between our URVSM and SOTA fine-tuned (unsupervised domain adaptation and supervised) methods on \textbf{Non Color Fundus} modality (excluding OCTA). The green boxes indicate the best result and the blue boxes indicate the second-best result.}
  \vspace{-0.1cm}
  \label{tab:main_tab_nonCF}
  \centering
  \scriptsize

  \begin{tabular}{@{\hspace{2pt}}c@{\hspace{2pt}}|cc|cccccccccc@{}}
  \toprule
     Dataset & \multicolumn{2}{c|}{Method} & Acc.$\uparrow$ & Dice$\uparrow$ & Sp$\uparrow$ & Se$\uparrow$ & Pr$\uparrow$ & F1$\uparrow$ & MCC$\uparrow$ & clDice$\uparrow$ & \(\beta_{0}\downarrow\) & \(\beta_{1}\downarrow\) \\
  \midrule
  \multirow{11}{*}{\shortstack{\ \(IOSTAR\) \ \\ (MC)}} & \multirow{4}{*}{\shortstack{\ \(Domain\) \ \\ \ \(Adaptation\) \ }} & Pse. Lab. & 96.27 & 74.53 & 98.42 & 71.10 & 79.35 & 74.53 & 72.92 & 80.55 & 75.3 & 9.1 \\
  & & FRR-TSNet & 96.25 & 73.37 & 98.72 & 67.39 & 81.93 & 73.37 & 72.11 & 79.11 & 73.1 & 10.1 \\
  & & MLAL & 95.33 & 72.85 & 96.57 & 80.85 & 67.14 & 72.85 & 70.99 & 77.61 & 180.1 & 9.8 \\
  & & MLAL-PDSF & 95.73 & 74.58 & 96.98 & 81.12 & 69.37 & 74.58 & 72.65 & 80.04 & 138.3 & \cellcolor{lightblue}{\textcolor{black}{6.3}} \\
  \cmidrule(l{0.1em}r{0.1em}){2-13}
  & \multirow{6}{*}{\(Supervised\)} & W-Net & \cellcolor{lightblue}{\textcolor{black}{97.09}} & \cellcolor{lightblue}{\textcolor{black}{80.66}} & \cellcolor{lightgreen}{\textcolor{black}{98.70}} & 78.54 & \cellcolor{lightgreen}{\textcolor{black}{83.50}} & \cellcolor{lightblue}{\textcolor{black}{80.66}} & \cellcolor{lightblue}{\textcolor{black}{79.30}} & \cellcolor{lightblue}{\textcolor{black}{87.46}} & \cellcolor{lightblue}{\textcolor{black}{43.2}} & 9.4 \\
  & & WS-DMF & 95.89 & 71.49 & 98.29 & 67.53 & 77.54 & 71.49 & 69.88 & 76.92 & 140.9 & 22.0 \\
  & & GT-DLA-dsHff & 96.91 & 79.63 & 98.52 & 78.27 & 81.67 & 79.63 & 78.16 & 86.99 & 43.4 & 7.8 \\
  & & U-Net & 97.01 & 80.50 & 98.52 & \cellcolor{lightblue}{\textcolor{black}{79.72}} & 81.80 & 80.50 & 79.04 & 87.08 & 57.4 & 8.7 \\
  & & CS$^{2}$-Net & 97.05 & 80.59 & \cellcolor{lightblue}{\textcolor{black}{98.59}} & 79.27 & \cellcolor{lightblue}{\textcolor{black}{82.49}} & 80.59 & 79.17 & 87.06 & 67.2 & 10.3 \\
  & & FA-Net & \cellcolor{lightgreen}{\textcolor{black}{97.13}} & \cellcolor{lightgreen}{\textcolor{black}{81.38}} & 98.52 & \cellcolor{lightgreen}{\textcolor{black}{81.30}} & 82.23 & \cellcolor{lightgreen}{\textcolor{black}{81.38}} & \cellcolor{lightgreen}{\textcolor{black}{80.06}} & \cellcolor{lightgreen}{\textcolor{black}{88.08}} & \cellcolor{lightgreen}{\textcolor{black}{38.3}} & 10.9 \\
  \cmidrule(l{0.1em}r{0.1em}){2-13}
  & \multicolumn{2}{c|}{\textbf{Our URVSM}} & 96.70 & 78.38 & 98.36 & 77.53 & 79.72 & 78.38 & 76.73 & 85.08 & 69.1 & \cellcolor{lightgreen}{\textcolor{black}{6.3}} \\

  \midrule

  \multirow{11}{*}{\shortstack{\ \(JRCFA\) \ \\ (FA)}} & \multirow{4}{*}{\shortstack{\ \(Domain\) \ \\ \ \(Adaptation\) \ }} & Pse. Lab. & 88.10 & 6.05 & 94.15 & 5.93 & 7.61 & 6.05 & 0.36 & 3.55 & 251.7 & 34.6 \\
  & & FRR-TSNet & 92.71 & 0.60 & 99.56 & 0.33 & 6.01 & 6.04 & -0.29 & 0.38 & 53.0 & 30.4 \\
  & & MLAL & 85.25 & 5.42 & 91.07 & 6.25 & 5.45 & 5.42 & -2.07 & 3.02 & 450.1 & 32.2 \\
  & & MLAL-PDSF & 86.47 & 4.86 & 92.49 & 5.25 & 5.43 & 4.86 & -1.88 & 2.73 & 411.6 & 23.5 \\
  \cmidrule(l{0.1em}r{0.1em}){2-13}
  & \multirow{6}{*}{\(Supervised\)} & W-Net & 97.83 & 83.49 & \cellcolor{lightblue}{\textcolor{black}{99.03}} & 81.73 & \cellcolor{lightgreen}{\textcolor{black}{86.11}} & 83.49 & 82.57 & 84.68 & 162.0 & 13.7 \\
  & & WS-DMF & 97.20 & 78.20 & 98.69 & 76.90 & 81.84 & 78.20 & 77.35 & 78.60 & 209.2 & 14.8 \\
  & & GT-DLA-dsHff & 96.38 & 73.14 & 98.17 & 72.25 & 74.44 & 73.14 & 71.31 & 72.16 & 279.1 & 13.2 \\
  & & U-Net & 97.89 & 83.95 & 99.14 & 81.65 & 87.02 & 83.95 & 83.04 & 84.86 & 160.5 & 13.9 \\
  & & CS$^{2}$-Net & \cellcolor{lightgreen}{\textcolor{black}{97.98}} & \cellcolor{lightgreen}{\textcolor{black}{84.88}} & 99.00 & \cellcolor{lightgreen}{\textcolor{black}{84.43}} & 85.79 & \cellcolor{lightgreen}{\textcolor{black}{84.88}} & \cellcolor{lightgreen}{\textcolor{black}{83.93}} & \cellcolor{lightgreen}{\textcolor{black}{86.07}} & 157.6 & \cellcolor{lightblue}{\textcolor{black}{11.2}} \\
  & & FA-Net & \cellcolor{lightblue}{\textcolor{black}{97.94}} & \cellcolor{lightblue}{\textcolor{black}{84.47}} & \cellcolor{lightgreen}{\textcolor{black}{99.03}} & \cellcolor{lightblue}{\textcolor{black}{83.51}} & \cellcolor{lightblue}{\textcolor{black}{85.94}} & \cellcolor{lightblue}{\textcolor{black}{84.47}} & \cellcolor{lightblue}{\textcolor{black}{83.51}} & \cellcolor{lightblue}{\textcolor{black}{85.72}} & \cellcolor{lightblue}{\textcolor{black}{126.3}} & \cellcolor{lightgreen}{\textcolor{black}{10.1}} \\
  \cmidrule(l{0.1em}r{0.1em}){2-13}
  & \multicolumn{2}{c|}{\textbf{Our URVSM}} & 96.47 & 75.00 & 97.91 & 78.40 & 72.89 & 74.89 & 73.43 & 76.37 & \cellcolor{lightgreen}{\textcolor{black}{37.7}} & 15.2 \\

  \midrule

  \multirow{11}{*}{\shortstack{\ \(JRCFAF\) \ \\ (FAF)}} & \multirow{4}{*}{\shortstack{\ \(Domain\) \ \\ \ \(Adaptation\) \ }} & Pse. Lab. & 94.26 & 63.03 & 94.76 & 84.39 & 52.93 & 63.03 & 63.34 & 66.55 & 438.9 & 30.3 \\
  & & FRR-TSNet & 97.52 & 74.34 & 98.73 & 73.91 & 76.60 & 74.34 & 73.53 & 76.86 & 136.5 & 7.6 \\
  & & MLAL & 95.21 & 64.64 & 95.69 & 85.59 & 53.70 & 64.64 & 64.99 & 64.45 & 482.7 & 10.9 \\
  & & MLAL-PDSF & 95.66 & 66.66 & 96.21 & 85.09 & 56.58 & 66.66 & 66.75 & 67.13 & 425.4 & 19.8 \\
  \cmidrule(l{0.1em}r{0.1em}){2-13}
  & \multirow{6}{*}{\(Supervised\)} & W-Net & 97.93 & 77.78 & \cellcolor{lightgreen}{\textcolor{black}{99.21}} & 74.01 & \cellcolor{lightgreen}{\textcolor{black}{82.59}} & 77.77 & 76.98 & 81.36 & 48.0 & 8.1  \\
  & & WS-DMF & 97.48 & 72.19 & 99.05 & 67.88 & 78.94 & 72.19 & 71.53 & 75.63 & 109.1 & 10.6 \\
  & & GT-DLA-dsHff & 97.24 & 70.71 & 98.80 & 67.71 & 74.57 & 70.71 & 69.50 & 74.61 & 62.8 & 7.5 \\
  & & U-Net & 97.89 & 77.66 & 99.13 & 74.94 & 81.29 & 77.66 & 76.81 & 81.03 & 68.6 & 7.2 \\
  & & CS$^{2}$-Net & \cellcolor{lightblue}{\textcolor{black}{97.93}} & \cellcolor{lightblue}{\textcolor{black}{78.11}} & \cellcolor{lightblue}{\textcolor{black}{99.14}} & 75.34 & \cellcolor{lightblue}{\textcolor{black}{81.64}} & \cellcolor{lightblue}{\textcolor{black}{78.12}} & \cellcolor{lightblue}{\textcolor{black}{77.24}} & 81.46 & 63.4 & \cellcolor{lightblue}{\textcolor{black}{6.9}} \\
  & & FA-Net & \cellcolor{lightgreen}{\textcolor{black}{97.95}} & \cellcolor{lightgreen}{\textcolor{black}{78.59}} & 99.09 & \cellcolor{lightblue}{\textcolor{black}{76.69}} & 81.17 & \cellcolor{lightgreen}{\textcolor{black}{78.69}} & \cellcolor{lightgreen}{\textcolor{black}{77.70}} & \cellcolor{lightgreen}{\textcolor{black}{82.55}} & \cellcolor{lightblue}{\textcolor{black}{42.4}} & 7.3 \\
  \cmidrule(l{0.1em}r{0.1em}){2-13}
  & \multicolumn{2}{c|}{\textbf{Our URVSM}} & 97.30 & 75.49 & 97.96 & \cellcolor{lightgreen}{\textcolor{black}{85.35}} & 68.19 & 75.49 & 74.78 & \cellcolor{lightblue}{\textcolor{black}{81.89}} & \cellcolor{lightgreen}{\textcolor{black}{21.6}} & \cellcolor{lightgreen}{\textcolor{black}{4.5}} \\

    \midrule

  \multirow{11}{*}{\shortstack{\ \(JRCIR\) \ \\ (IR)}} & \multirow{4}{*}{\shortstack{\ \(Domain\) \ \\ \ \(Adaptation\) \ }} & Pse. Lab. & 95.91 & 71.89 & 97.84 & 71.85 & 73.17 & 71.89 & 70.05 & 71.15 & 86.9 & 8.3 \\
  & & FRR-TSNet & 96.06 & 71.95 & 98.21 & 69.38 & 76.13 & 71.95 & 70.30 & 71.02 & 112.1 & 15.3 \\
  & & MLAL & 94.84 & 69.21 & 96.26 & 76.85 & 64.93 & 69.21 & 67.44 & 68.55 & 239.4 & 14.1 \\
  & & MLAL-PDSF & 95.31 & 70.07 & 97.00 & 74.25 & 68.91 & 70.07 & 68.43 & 69.12 & 205.6 & 11.2 \\
  \cmidrule(l{0.1em}r{0.1em}){2-13}
  & \multirow{6}{*}{\(Supervised\)} & W-Net & 97.07 & 79.79 & \cellcolor{lightgreen}{\textcolor{black}{98.85}} & 75.25 & \cellcolor{lightgreen}{\textcolor{black}{83.35}} & 78.80 & 77.52 & 79.19 & 55.3 & 7.0 \\
  & & WS-DMF & 96.32 & 74.49 & 98.10 & 74.37 & 75.87 & 74.49 & 72.88 & 75.67 & 107.9 & 9.8 \\
  & & GT-DLA-dsHff & 96.30 & 73.68 & 98.30 & 71.36 & 76.64 & 73.68 & 71.88 & 74.46 & 88.5 & 5.9 \\
  & & U-Net & 97.14 & 79.92 & 98.66 & \cellcolor{lightblue}{\textcolor{black}{78.74}} & 81.84 & 79.92 & 78.60 & 80.93 & 79.4 & 5.6 \\
  & & CS$^{2}$-Net & \cellcolor{lightblue}{\textcolor{black}{97.22}}
 & \cellcolor{lightblue}{\textcolor{black}{80.28}} & \cellcolor{lightblue}{\textcolor{black}{98.77}} & 78.30 & \cellcolor{lightblue}{\textcolor{black}{83.01}} & \cellcolor{lightblue}{\textcolor{black}{80.28}} & \cellcolor{lightblue}{\textcolor{black}{79.01}} & \cellcolor{lightblue}{\textcolor{black}{80.79}} & 86.9 & 5.6 \\
  & & FA-Net & \cellcolor{lightgreen}{\textcolor{black}{97.24}} & \cellcolor{lightgreen}{\textcolor{black}{80.48}} & 98.75 & \cellcolor{lightgreen}{\textcolor{black}{78.79}} & 82.84 & \cellcolor{lightgreen}{\textcolor{black}{80.48}} & \cellcolor{lightgreen}{\textcolor{black}{79.19}} & \cellcolor{lightgreen}{\textcolor{black}{81.30}} & \cellcolor{lightblue}{\textcolor{black}{49.6}} & \cellcolor{lightgreen}{\textcolor{black}{5.0}} \\
  \cmidrule(l{0.1em}r{0.1em}){2-13}
  & \multicolumn{2}{c|}{\textbf{Our URVSM}} & 96.10 & 74.66 & 97.84 & 75.31 & 72.85 & 73.66 & 71.81 & 76.30 & \cellcolor{lightgreen}{\textcolor{black}{28.8}} & \cellcolor{lightblue}{\textcolor{black}{5.5}} \\

  \bottomrule
  \end{tabular}
  \vspace{-0.5cm}
\end{table*}

\begin{table*}[ht]
  \caption{Comparison between our URVSM with SOTA fine-tuned (unsupervised domain adaptation and supervised) methods on \textbf{Color Fundus} modality. The green boxes indicate the best result and the blue boxes indicate the second-best result.}
  \vspace{-0.1cm}
  \label{tab:main_tab_CF}
  \centering
  \scriptsize

  \begin{tabular}{@{\hspace{2pt}}c@{\hspace{2pt}}|cc|cccccccccc@{}}
  \toprule
     Dataset & \multicolumn{2}{c|}{Method} & Acc.$\uparrow$ & Dice$\uparrow$ & Sp$\uparrow$ & Se$\uparrow$ & Pr$\uparrow$ & F1$\uparrow$ & MCC$\uparrow$ & clDice$\uparrow$ & \(\beta_{0}\downarrow\) & \(\beta_{1}\downarrow\) \\
  \midrule
  \multirow{11}{*}{\ \(STARE\) \ } & \multirow{4}{*}{\shortstack{\ \(Domain\) \ \\ \ \(Adaptation\) \ }} & Pse. Lab. & 97.17 & 76.06 & \cellcolor{lightblue}{\textcolor{black}{98.99}} & 70.97 & \cellcolor{lightblue}{\textcolor{black}{84.07}} & 76.06 & 75.36 & 80.71 & 71.7 & 19.0 \\
  & & FRR-TSNet & 97.15 & 74.41 & \cellcolor{lightgreen}{\textcolor{black}{99.10}} & 68.72 & \cellcolor{lightgreen}{\textcolor{black}{85.88}} & 74.41 & 74.44 & 79.31 & \cellcolor{lightblue}{\textcolor{black}{64.1}} & 19.4 \\
  & & MLAL & 95.85 & 69.96 & 97.17 & 76.01 & 68.06 & 69.96 & 68.95 & 73.30 & 122.1 & 15.3 \\
  & & MLAL-PDSF & 96.39 & 73.12 & 97.76 & 76.73 & 71.89 & 73.12 & 71.88 & 76.09 & 129.2 & 15.6 \\
  \cmidrule(l{0.1em}r{0.1em}){2-13}
  & \multirow{6}{*}{\(Supervised\)} & W-Net & 97.61 & 81.52 & 98.80 & 81.59 & 82.36 & 81.52 & 80.51 & 85.56 & 75.6 & 13.1 \\
  & & WS-DMF & 96.01 & 71.77 & 97.36 & 77.28 & 68.55 & 71.77 & 70.30 & 78.94 & 193.3 & 17.3 \\
  & & GT-DLA-dsHff & 95.59 & 66.17 & 97.70 & 66.15 & 66.71 & 66.17 & 63.96 & 66.44 & 84.0 & 12.2 \\
  & & U-Net & 97.25 & 78.32 & 98.67 & 77.27 & 80.48 & 78.32 & 77.17 & 81.59 & 132.7 & 14.3 \\
  & & CS$^{2}$-Net & \cellcolor{lightblue}{\textcolor{black}{97.80}} & \cellcolor{lightblue}{\textcolor{black}{82.86}} & 98.91 & \cellcolor{lightblue}{\textcolor{black}{82.85}} & 83.85 & \cellcolor{lightblue}{\textcolor{black}{82.86}} & \cellcolor{lightblue}{\textcolor{black}{81.97}} & \cellcolor{lightblue}{\textcolor{black}{86.41}} & 84.9 & 12.8 \\
  & & FA-Net & \cellcolor{lightgreen}{\textcolor{black}{97.85}}  & \cellcolor{lightgreen}{\textcolor{black}{83.57}} & 98.84 & \cellcolor{lightgreen}{\textcolor{black}{84.51}} & 83.17 & \cellcolor{lightgreen}{\textcolor{black}{83.57}} & \cellcolor{lightgreen}{\textcolor{black}{82.58}} & \cellcolor{lightgreen}{\textcolor{black}{87.63}} & 69.1 & \cellcolor{lightgreen}{\textcolor{black}{10.1}} \\
  \cmidrule(l{0.1em}r{0.1em}){2-13}
  & \multicolumn{2}{c|}{\textbf{Our URVSM}} & 97.48 & 80.51 & 98.69 & 80.77 & 80.78 & 80.52 & 79.31 & 85.21 & \cellcolor{lightgreen}{\textcolor{black}{35.1}} & \cellcolor{lightblue}{\textcolor{black}{12.6}} \\

  \midrule

  \multirow{11}{*}{\ \(HRF\) \ } & \multirow{4}{*}{\shortstack{\ \(Domain\) \ \\ \ \(Adaptation\) \ }} & Pse. Lab. & \cellcolor{lightblue}{\textcolor{black}{97.00}} & \cellcolor{lightblue}{\textcolor{black}{67.36}} & \cellcolor{lightblue}{\textcolor{black}{98.83}} & 62.73 & 74.10 & \cellcolor{lightblue}{\textcolor{black}{67.36}} & \cellcolor{lightblue}{\textcolor{black}{66.38}} & 60.19 & \cellcolor{lightgreen}{\textcolor{black}{69.6}} & 71.5 \\
  & & FRR-TSNet & \cellcolor{lightgreen}{\textcolor{black}{97.07}} & 66.27 & \cellcolor{lightgreen}{\textcolor{black}{99.12}} & 58.55 & 78.75 & 66.27 & 66.03 & 59.22 & \cellcolor{lightblue}{\textcolor{black}{75.1}} & 74.1 \\
  & & MLAL & 95.65 & 61.96 & 96.94 & 71.07 & 55.87 & 61.96 & 60.52 & 60.77 & 103.8 & 59.0 \\
  & & MLAL-PDSF & 95.93 & 63.37 & 97.26 & 70.95 & 57.89 & 63.37 & 61.81 & 61.71 & 95.0 & 60.3 \\
  \cmidrule(l{0.1em}r{0.1em}){2-13}
  & \multirow{6}{*}{\(Supervised\)} & W-Net & 96.60 & 64.39 & 98.48 & 61.70 & \cellcolor{lightblue}{\textcolor{black}{68.04}} & 64.39 & 62.88 & 60.09 & 260.4 & 59.8 \\
  & & WS-DMF & 94.79 & 58.02 & 96.02 & \cellcolor{lightblue}{\textcolor{black}{71.78}} & 49.75 & 58.03 & 56.84 & 56.54 & 945.1 & \cellcolor{lightblue}{\textcolor{black}{51.7}} \\
  & & GT-DLA-dsHff & 95.49 & 53.17 & 97.86 & 51.16 & 55.84 & 53.17 & 50.99 & 45.92 & 421.0 & 60.0 \\
  & & U-Net & 96.58 & 64.22 & 98.42 & 61.82 & 67.40 & 64.22 & 62.64 & 59.63 & 297.7 & 53.1 \\
  & & CS$^{2}$-Net & 96.68 & 65.17 & 98.51 & 62.37 & \cellcolor{lightgreen}{\textcolor{black}{68.82}} & 65.18 & 63.67 & 60.82 & 334.7 & \cellcolor{lightgreen}{\textcolor{black}{51.1}} \\
  & & FA-Net & 96.67 & 65.80 & 98.40 & 64.36 & 67.87 & 65.80 & 64.23 & \cellcolor{lightblue}{\textcolor{black}{61.97}} & 257.7 & 47.5 \\
  \cmidrule(l{0.1em}r{0.1em}){2-13}
  & \multicolumn{2}{c|}{\textbf{Our URVSM}} & 96.60 & \cellcolor{lightgreen}{\textcolor{black}{68.50}} & 97.78 & \cellcolor{lightgreen}{\textcolor{black}{74.70}} & 63.87 & \cellcolor{lightgreen}{\textcolor{black}{68.50}} & \cellcolor{lightgreen}{\textcolor{black}{67.15}} & \cellcolor{lightgreen}{\textcolor{black}{68.26}} & 142.8 & 55.5 \\

  \midrule

  \multirow{11}{*}{\ \(ChaseDB1\) \ } & \multirow{4}{*}{\shortstack{\ \(Domain\) \ \\ \ \(Adaptation\) \ }} & Pse. Lab. & 96.07 & 71.24 & 98.25 & 68.35 & 75.18 & 71.25 & 69.44 & 70.05 & 68.6 & \cellcolor{lightgreen}{\textcolor{black}{6.8}} \\
  & & FRR-TSNet & 95.94 & 68.93 & 98.53 & 62.93 & 76.63 & 68.93 & 67.26 & 67.17 & 83.5 & 18.3 \\
  & & MLAL & 94.99 & 68.96 & 96.30 & 78.01 & 62.07 & 68.96 & 66.89 & 67.44 & 203.1 & 23.2 \\
  & & MLAL-PDSF & 95.59 & 71.17 & 97.06 & 76.61 & 66.73 & 71.17 & 69.08 & 69.93 & 146.3 & 18.1 \\
  \cmidrule(l{0.1em}r{0.1em}){2-13}
  & \multirow{6}{*}{\(Supervised\)} & W-Net & 97.08 & 78.87 & \cellcolor{lightblue}{\textcolor{black}{98.75}} & 76.35 & \cellcolor{lightgreen}{\textcolor{black}{82.32}} & 78.87 & 77.58 & 78.46 & \cellcolor{lightgreen}{\textcolor{black}{54.09}} & 9.0 \\
  & & WS-DMF & 95.54 & 59.62 & \cellcolor{lightgreen}{\textcolor{black}{98.81}} & 50.31 & 78.94 & 59.62 & 59.92 & 58.15 & 123.6 & 24.1 \\
  & & GT-DLA-dsHff & 96.40 & 74.26 & 98.24 & 72.87 & 76.11 & 74.26 & 72.46 & 73.06 & 81.1 & 7.6 \\
  & & U-Net & 96.81 & 76.84 & 98.60 & 74.23 & 80.36 & 76.84 & 75.39 & 76.01 & 83.7 & 8.2 \\
  & & CS$^{2}$-Net & \cellcolor{lightgreen}{\textcolor{black}{97.19}} & \cellcolor{lightblue}{\textcolor{black}{80.05}} & 98.65 & \cellcolor{lightblue}{\textcolor{black}{78.89}} & \cellcolor{lightblue}{\textcolor{black}{81.65}} & \cellcolor{lightblue}{\textcolor{black}{80.05}} & \cellcolor{lightblue}{\textcolor{black}{78.67}} & \cellcolor{lightblue}{\textcolor{black}{80.77}} & 84.9 & 8.6 \\
  & & FA-Net & \cellcolor{lightblue}{\textcolor{black}{97.18}} & \cellcolor{lightgreen}{\textcolor{black}{80.21}} & 98.52 & \cellcolor{lightgreen}{\textcolor{black}{80.39}} & 80.72 & \cellcolor{lightgreen}{\textcolor{black}{80.21}} & \cellcolor{lightgreen}{\textcolor{black}{78.89}} & \cellcolor{lightgreen}{\textcolor{black}{81.29}} & \cellcolor{lightblue}{\textcolor{black}{68.4}} & \cellcolor{lightblue}{\textcolor{black}{7.1}} \\
  \cmidrule(l{0.1em}r{0.1em}){2-13}
  & \multicolumn{2}{c|}{\textbf{Our URVSM}} & 96.55 & 77.34 & 98.05 & 77.81 & 75.38 & 76.34 & 74.63 & 78.34 & 70.1 & 7.5 \\

  \bottomrule
  \end{tabular}
  \vspace{-0.5cm}
\end{table*}

\begin{table}
  \caption{Performance on Ultra-Wide-Field retinal images, pure zero-shot performance on OCTA, and comparison with SOTA supervised methods.}
  \vspace{-0.1cm}
  \label{tab:OCTA_UWF}
  \centering
  \scriptsize

  \begin{tabular}{@{\hspace{2pt}}c@{\hspace{2pt}}|c|cccc@{}}
  \toprule
     Dataset & Method & Acc$\uparrow$ & Dice$\uparrow$ & Sp$\uparrow$ & Se$\uparrow$ \\
     
  \midrule
  \multirow{5}{*}{\shortstack{\(OCTA500-3M\) \\ (OCTA)}} & UNet++ & 95.98 & 88.64 & - & - \\
   & FARGO & 98.12 & 91.68 & - & - \\
   & FRNet-base & 98.95 & 91.29 & 99.50 & 90.56 \\
   & FRNet & \textbf{99.00} & \textbf{91.70} & \textbf{99.51} & \textbf{91.19} \\
   & \textbf{Our URVSM} & 96.29 & 74.68 & 97.36 & 81.97 \\

  \midrule
  \multirow{5}{*}{\shortstack{\(OCTA500-6M\) \\ (OCTA)}} & UNet++ & 95.73 & 85.76 & - & - \\
   & FARGO & 98.12 & \textbf{89.15} & - & - \\
   & FRNet-base & 98.11 & 88.82 & \textbf{99.05} & 88.01 \\
   & FRNet & \textbf{98.14} & 89.03 & 99.01 & \textbf{88.77} \\
   & \textbf{Our URVSM} & 95.21 & 71.68 & 98.20 & 66.53 \\

  \toprule
  Dataset & Method & Dice$\uparrow$ & IoU$\uparrow$ & MCC$\uparrow$ & BM$\uparrow$ \\
  
  \midrule
  \multirow{5}{*}{\shortstack{\(PRIME-FP20\) \\ (UWF-MC)}} & LKM-UNet & 58.11 & 41.00 & 57.55 & 51.55 \\
   & Swin-Umamba & 64.99 & 48.43 & 64.36 & 61.93 \\
   & EM-Net & 62.30 & 45.31 & 61.84 & 55.40 \\
   & Serp-Mamba & \textbf{69.01} & \textbf{52.93} & \textbf{68.41} & 65.75 \\
   & \textbf{Our URVSM} & 65.40 & 49.00 & 65.69 & \textbf{74.32} \\

  \bottomrule
  \end{tabular}
  \vspace{-0.5cm}
\end{table}

\begin{table}
  \caption{Comparison between URVSM and general-purpose foundational segmentation models, metric reported: Dice score.}
  \vspace{-0.1cm}
  \label{tab:sam_comp}
  \centering
  \scriptsize

  \begin{tabular}{@{\hspace{2pt}}c@{\hspace{2pt}}|cccccc@{}}
  \toprule
     \ & CF & MC & FA & FAF & IR & OCTA \\
  \midrule
  \multicolumn{1}{c|}{SAM2} & 15.45 & 18.19 & 16.12 & 12.05 & 17.08 & 15.80 \\
  \multicolumn{1}{c|}{Grounded SAM2} & 17.57 & 18.11 & 15.77 & 12.04 & 17.10 & 16.34 \\
  \multicolumn{1}{c|}{MedSAM2} & 14.74 & 16.99 & 9.68 & 6.90 & 6.66 & 14.58 \\
  \multicolumn{1}{c|}{\textbf{Our URVSM}} & \textbf{76.21} & \textbf{78.38} & \textbf{75.00} & \textbf{75.49} & \textbf{74.66} & \textbf{72.54} \\

  \bottomrule
  \end{tabular}
\end{table}

\subsection{Datasets}
\subsubsection{Image Translation Training}
\label{sssec:trans_dataset}
We prepared a multi-modality database with over 150k de-identified images from the database of the Jacobs Retina Center at UCSD. However, in our previous numerous attempts we found it very difficult and time-consuming to tune the training of the image translation network (regardless of the methods) for a large dataset. Therefore, in this work, we use a 4k-image (2k CF, 500 FA, FAF, MC, IR/NIR, respectively) dataset from the larger 150k database. Each image has a resolution of 768 $\times$ 768. All the CF images are from the 30-degree (field-of-view) Topcon camera and all the IR/NIR images are from the Heidelberg camera. For FA, FAF and MC, 250 images are from the Heidelberg camera. For the Optos camera, we crop random regions from the original 135 degrees to get 30-40-degree conventional field-of-view images (so that the image size is the same as the other modalities) and apply a commonly used circular mask (\cref{fig:masks}(a)). In the dataset, there are 250 cropped and masked Optos images for FA, MC, FAF modalities, respectively.

\subsubsection{Segmentation Training}
\label{sssec:segtrain_dataset}
We train the segmentation network using only the 20 training images from the public DRIVE \cite{Dataset_DRIVE} CF dataset. In practice, we discard image \#14 due to the unusual vessel pattern. In addition, for all the images we resize them to 768$\times$768 for training. We also resized the ground truth annotations and manually corrected the errors caused by interpolation. In addition, according to our examination, the original DRIVE ground truth has multiple topological errors such as small redundant holes in the vessel foreground. In our revised ground truth, we correct these errors and allow more efficient topological learning.

\subsubsection{Evaluation}
Ten diverse datasets were used for evaluation. Their details can be found in \cref{tab:datasets_details}. For the CF modality, we use the public STARE \cite{Dataset-STARE}, ChaseDB1 \cite{Dataset-ChaseDB1}, and HRF \cite{Dataset-HRF} datasets. For MC modality, we use the public IOSTAR \cite{Dataset-IOSTAR} dataset. For FA, FAF and IR/NIR, we prepared three new datasets: JRCFA, JRCFAF, and JRCIR. Each dataset has 40 768$\times$768 images, which is more than most of the commonly used public retinal vessel segmentation (CF) datasets. The JRCFA and JRCFAF datasets each consist of 20 images from Heidelberg and 20 from Optos, respectively. The JRCIR dataset consists of 40 Heidelberg (the only camera for this modality) images. All the vessel annotations are either labeled or carefully examined and refined by the retinal experts in the UCSD Jacobs Retina Center (JRC). Furthermore, we use an OCTA500-3M and an OCTA600-6M dataset \cite{Dataset-OCTA500} to evaluate the performance of the URVSM on OCTA in two different fields of view (3mm and 6mm). For the special UWF images (technically UWF refers to field-of-view rather than a distinct modality), we use the PRIME-FP20 (MC) dataset \cite{Dataset-PRIMEFP20}. In addition to the ten vessel segmentation datasets, we also evaluate our model qualitatively on the Optos MC images using images from our 150K database and on the Canon FA images from the CF-FA registration dataset \cite{Dataset-CFFA}. We provide preliminary insights in this paper (\cref{fig:intro}, \cref{fig:main_quali}), yet full evaluation requires the vessel labels (unavailable) and is left as future work. Additionally, we use the Massachusetts Roads \cite{data_road} (road segmentation in aerial imagery) and CREMI \cite{data_cremi} (extracellular boundary segmentation in electronic microscopy) datasets for the extended experiments on topological domain adaptation on more general curvilinear structure segmentation. The test set of Massachusetts Roads and all slices in volume C in CREMI were used for evaluation.

\subsection{Implementation Details}
\subsubsection{Translation Network}
The balance between the size of the generator and discriminator is important to stably train the CycleGAN for our task. We use a U-Net \cite{Others-U-Net} generator with 16.66M parameters and an NLP discriminator with 2.77M parameters\footnote{More details on the network structures we use for translation and segmentation are provided in our code.}. For training, we use a batch size of 4 and train the model for 100 epochs on the 4000 images. We start with an initial learning rate of 2e-4 and start decaying the learning rate linearly to 0 from epoch 50 to 100. Adam optimizer is used to train both the generators and the discriminators, with a momentum of $(0.5, 0.999)$ and no weight decay. $\lambda_{c}=$10.0 and $\lambda_{i}=$0.5.

\subsubsection{Segmentation Network}
An initial learning rate of 1e-3 is used to train the model for 150 epochs, then decayed to 1e-4 to train for another 250 epochs. Adam optimizer is used with a weight decay of 1e-3 and momentum of $(0.9, 0.999)$. We train with a batch size of 1. The weights $\lambda_{tc}$ and $\lambda_{ts}$ are 0.05 and 0.0002, respectively. All the testing images are first padded to square and resized to 768$\times$768 to pass through the translation and the segmentation network. Then the prediction is resized back to the original image size to compute the metrics with the ground truth labels. The implementation of both image translation and the segmentation parts are under the PyTorch framework. The computation of persistent homology relies on the GUDHI library \cite{gudhi:urm}. The computational cost of this work is reported in \cref{tab:comp_cost}.

\subsubsection{Data Augmentation}
Extensive data augmentations are used for training both the translation and the segmentation network. For translation network, random scaling, sharpness, contrast and gamma adjustment are used. Additionally, since different camera manufacturers apply different shapes of mask to the image, we randomly apply the masks shown in \cref{fig:masks} to the images to prevent the model from generating artifacts out of the effective imaging area. For segmentation network, in addition to the aforementioned methods, we also apply random negative film followed by a contrast or sharpness adjustment. This is essential to allow the universal segmentation model to work for the FA modality as the vessels appear to be brighter than the background and we found no image translation method capable of stably inverting the color of the vessels. In addition, random image rotation, shift and Gaussian noise are applied.

\subsection{Comparison with Fine-tuned SOTA Methods}
\subsubsection{URVSM on retinal images in conventional field-of-view}
To the best of our knowledge, this work is the first universal retinal vessel segmentation model and there is no previous method that we can fairly compare with. Therefore, we compare with state-of-the-art (SOTA) methods which are fine-tuned individually on each dataset either by full-supervision or unsupervised domain adaptation. Since we use three new datasets and the works in the retinal vessel segmentation literature have inconsistent dataset choice, dataset split, and inconsistent choice of evaluation metrics, we prepare a new benchmark where all the methods are evaluated in a consistent way for fair comparison. The baseline methods are run using the recommended training details in respective papers and in a five-fold experiment that splits the whole dataset by 5 groups. Each fold uses one group for testing and the other four for training. Eventually, all images are used for testing so as to make sure that the average score is computed on the same images with our universal model (which is directly tested on all the images). For the unsupervised domain adaptation methods, we use the DRIVE dataset with annotation as the source domain and the per-fold training set of each dataset as the target domain for training. The supervised method baseline includes U-Net \cite{Others-U-Net}, CS$^{2}$-Net \cite{Vesseg-CNN-CS2Net}, W-Net \cite{Vesseg-CNN-WNet}, FA-Net \cite{Vesseg-CNN-FANet}, GT-DLA-dsHff \cite{Vesseg-CNN-GlobalTransformer} and WS-DMF \cite{Vesseg-CNN-WSDMF}. For domain adaptation, the baselines include Pse. Lab. (pseudo-label-based) \cite{Vesseg-CNN-WNet}, FRR-TSNet (teacher-student network-based) \cite{FRR-TSNet}, MLAL (adversarial learning-based) and MLAL-PDSF (hybrid adversarial learning and teacher-student) \cite{MLAL-PDSF}. The training and testing of the baselines are done in original image size except for the HRF dataset. The images in HRF dataset are too large to be trained on our 24GB GPUs. Therefore, we first pad the image and ground truth to square, then downsample them to 768$\times$768 for training and inference. The predicted vessel likelihood is then upsampled to the original image size to compare with the original ground truth and compute the metrics. 

Since the use of evaluation metrics is inconsistent in the literature, in our work we adopt a consistent benchmark and use most of the commonly used metrics, including the pixelwise accuracy (Acc.), the Dice score, the specificity (Sp), the sensitivity (Se), the Precision (Pr), the F1 score, and the Matthews Correlation Coefficients (MCC). In addition, although the main goal of topological feature learning in this work is for domain generalization, topological accuracy remains crucial for retinal vessel segmentation. Therefore, we also use three topological metrics which are commonly used in the segmentation of more general tubular structures. These metrics include the clDice score \cite{Topo-clDice} and the 0-/1-Betti errors ($\beta_{0}, \beta_{1}$) \cite{Topo-WTLoss}. For all of the reported metrics, we compute them at the original image size and average over all test images.

The results in \cref{tab:main_tab_nonCF}, \cref{tab:main_tab_CF} and \cref{fig:main_quali} show that our proposed model, not only allows universal segmentation, but also has comparable performance with SOTA fine-tuned supervised methods and outperforms the fine-tuned unsupervised domain adaptation methods. However, for some modalities there still exists non-negligible performance gap compared with the supervised methods, especially for the FA. As shown by an example in the third row of \cref{fig:main_quali}, the FA modality (particularly from the Optos camera) has very high contrast between the vessel and the background. As a result, more ending vanishing vessels are visible on the image and the universal model often fails to identify them. The reasons are two-fold: (1) The image translation introduces non-avoidable detail loss; (2) The topological loss function makes the model more conservative in predicting a pixel as vessel.

\begin{figure*}[ht]
\centering
\includegraphics[width=0.99\textwidth]{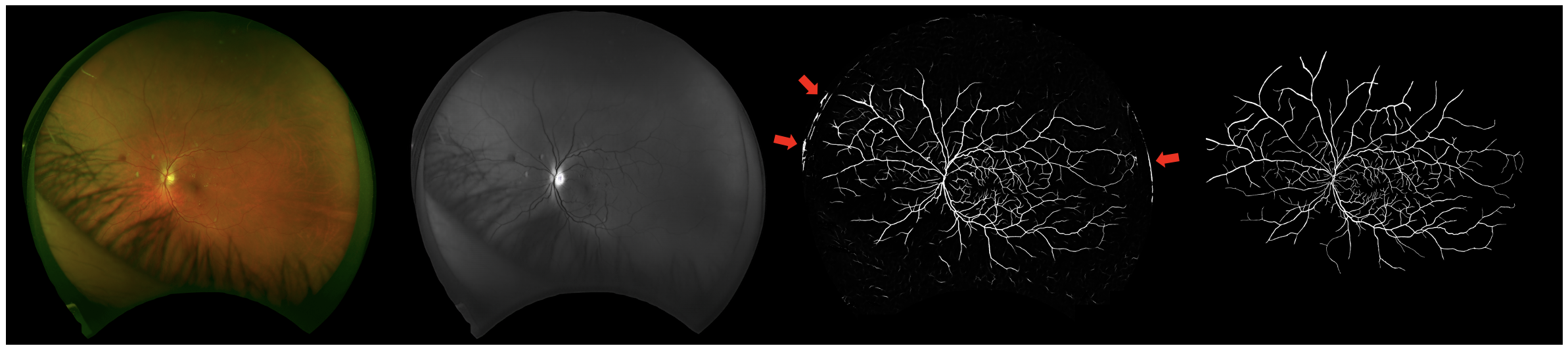}
\caption{Qualitative results on Ultra-Wide-Field (UWF) retinal images. From left to right: image, translated image, URVSM segmentation, ground truth. Red arrows show the outlier patterns in the peripheral causing false segmentation, which is a current limitation of the URVSM.}
\label{fig:UWF_quali}
\vspace{-0.2cm}
\end{figure*}

\begin{figure*}[ht]
\centering
\includegraphics[width=0.98\textwidth]{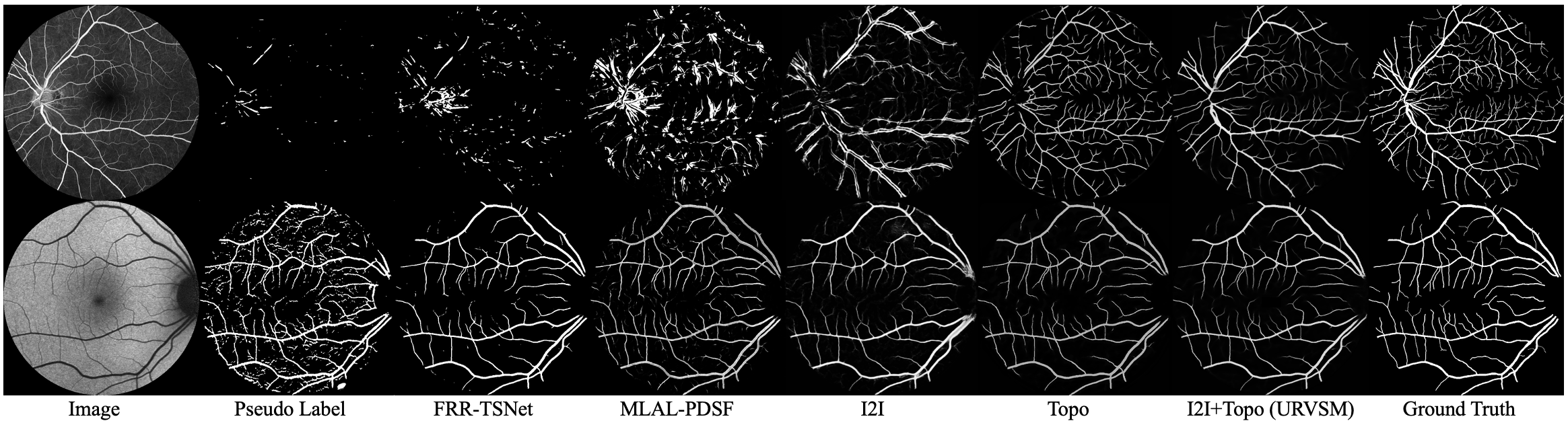}
\vspace{-0.3cm}
\caption{Qualitative result on how each component in the proposed pipeline acts and further comparison with other domain adaptation methods.}
\label{fig:DA_FA}
\vspace{-0.1cm}
\end{figure*}

\begin{figure}[]
\centering
\includegraphics[width=0.47\textwidth]{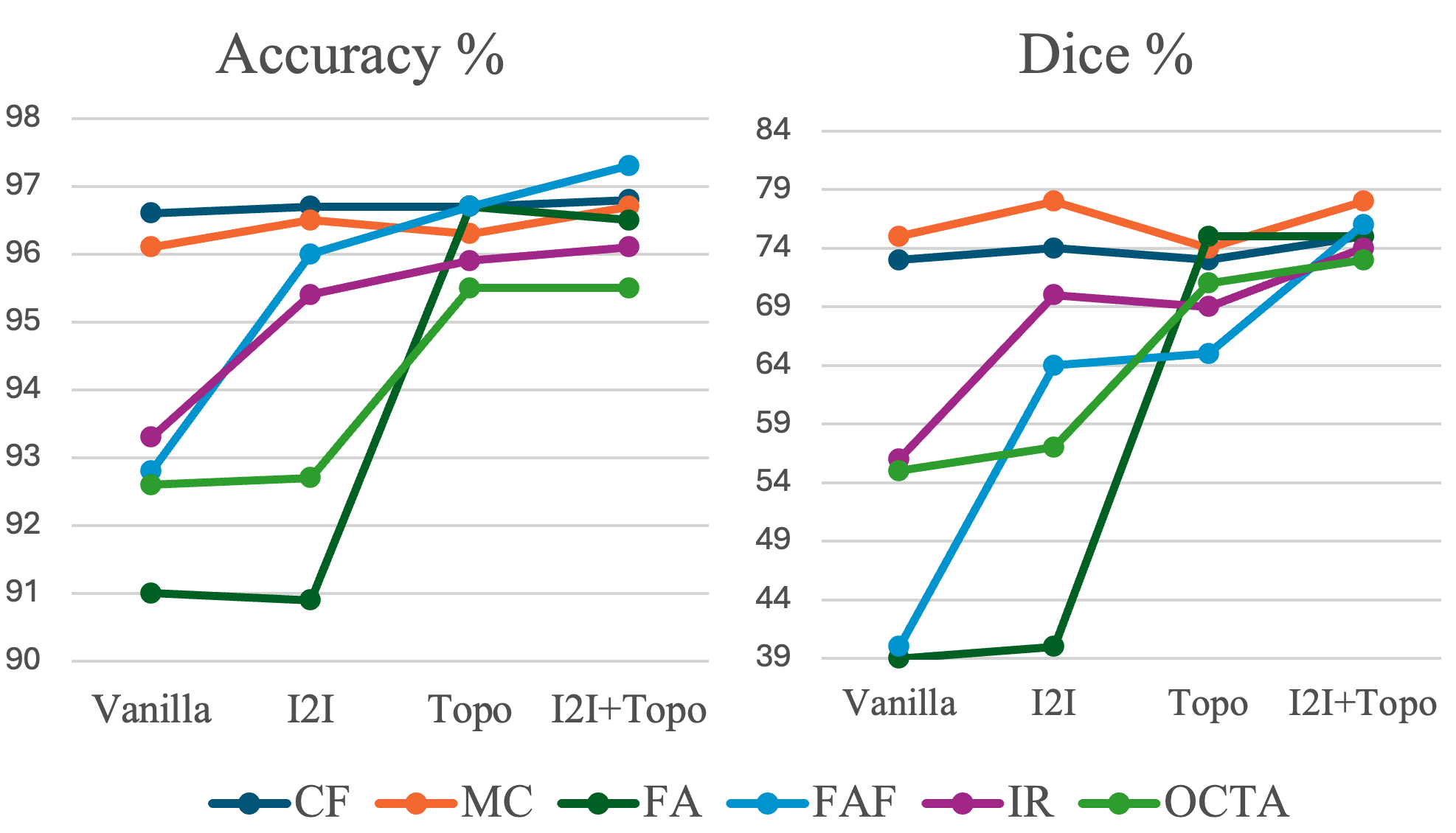}
\vspace{-0.3cm}
\caption{Ablation studies on how each proposed step in the pipeline improves the performance of the universal vessel segmentation model. `I2I' refers to image-to-image translation.}
\label{fig:ab_step}
\vspace{-0.4cm}
\end{figure}

\subsubsection{URVSM on OCTA and UWF}
We separately evaluate OCTA modality and Ultra-Wide-Field retinal images (UWF) since they have considerably different fields of view and the image features in the OCTA modality are significantly different from the other modalities. In addition, vessel segmentation of these two types of retinal images has its independent literature and established baselines. For the metrics, we follow and use the metrics in \cite{Dataset-OCTA500} for OCTA modality and the metrics in \cite{UWF_SerpMamba} for UWF imaging.

For OCTA, we compare with SOTA supervised methods: UNet++ \cite{UNet++}, FARGO \cite{FARGO}, FRNet-base and FRNet \cite{Dataset-OCTA500}. We observed that the URVSM has good performance in predicting the major vessels, whereas the prediction for the thin ending vessels is usually missing (\cref{fig:main_quali}). In general, there are still notable performance gaps between the URVSM and SOTA fine-tuned methods (\cref{tab:OCTA_UWF}). It should be noted that we only use 19 annotated CF images to train the model, while our baseline methods are fine-tuned on OCTA training set with highly similar data with the testing set and are over 7 times larger. Overall, the zero-shot results on OCTA show the strong generalization ability of our URVSM and the effectiveness of our proposed learning methods.

For UWF, we compare with SOTA supervised methods: LKM-UNet \cite{UWF_LKMUNet}, Swin-Umamba \cite{UWF_Swin-UMamba}, EM-Net \cite{UWF_EMNet} and Serp-Mamba \cite{UWF_SerpMamba}. Experiments show that the URVSM has strong performance in the center area of the retina but can generate more artifacts at the boundary of the image (\cref{fig:UWF_quali}). The overall performance is comparable with the state-of-the-art methods (\cref{tab:OCTA_UWF}). The results on UWF images again shows the strong generalization ability of our proposed model.


\begin{figure*}[ht]
\centering
\includegraphics[width=0.98\textwidth]{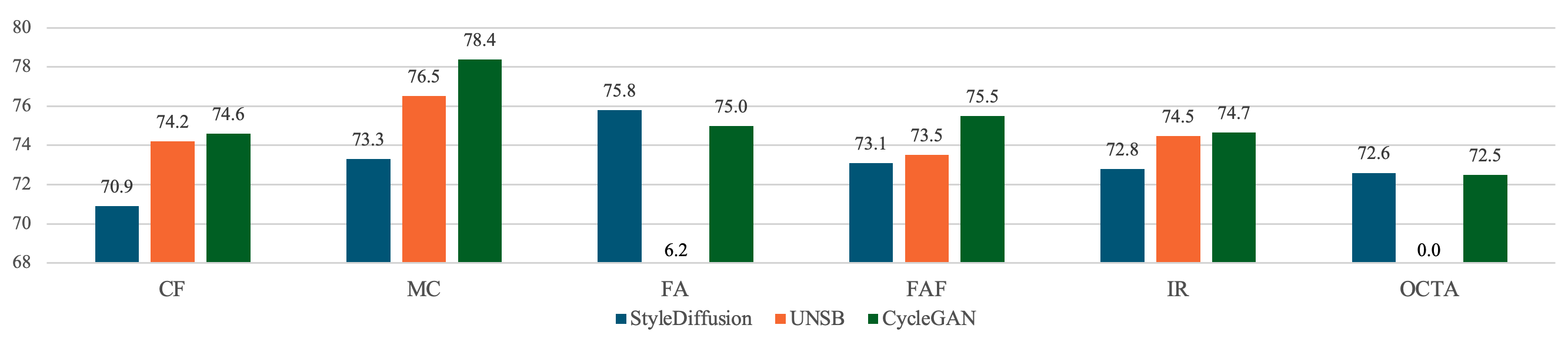}
\vspace{-0.3cm}
\caption{Comparison of the final segmentation performance between using different methods for image translation. Metric reported: Dice score.}
\label{fig:ab_transfer_chart}
\vspace{-0.2cm}
\end{figure*}

\begin{figure}[ht]
\centering
\includegraphics[width=0.48\textwidth]{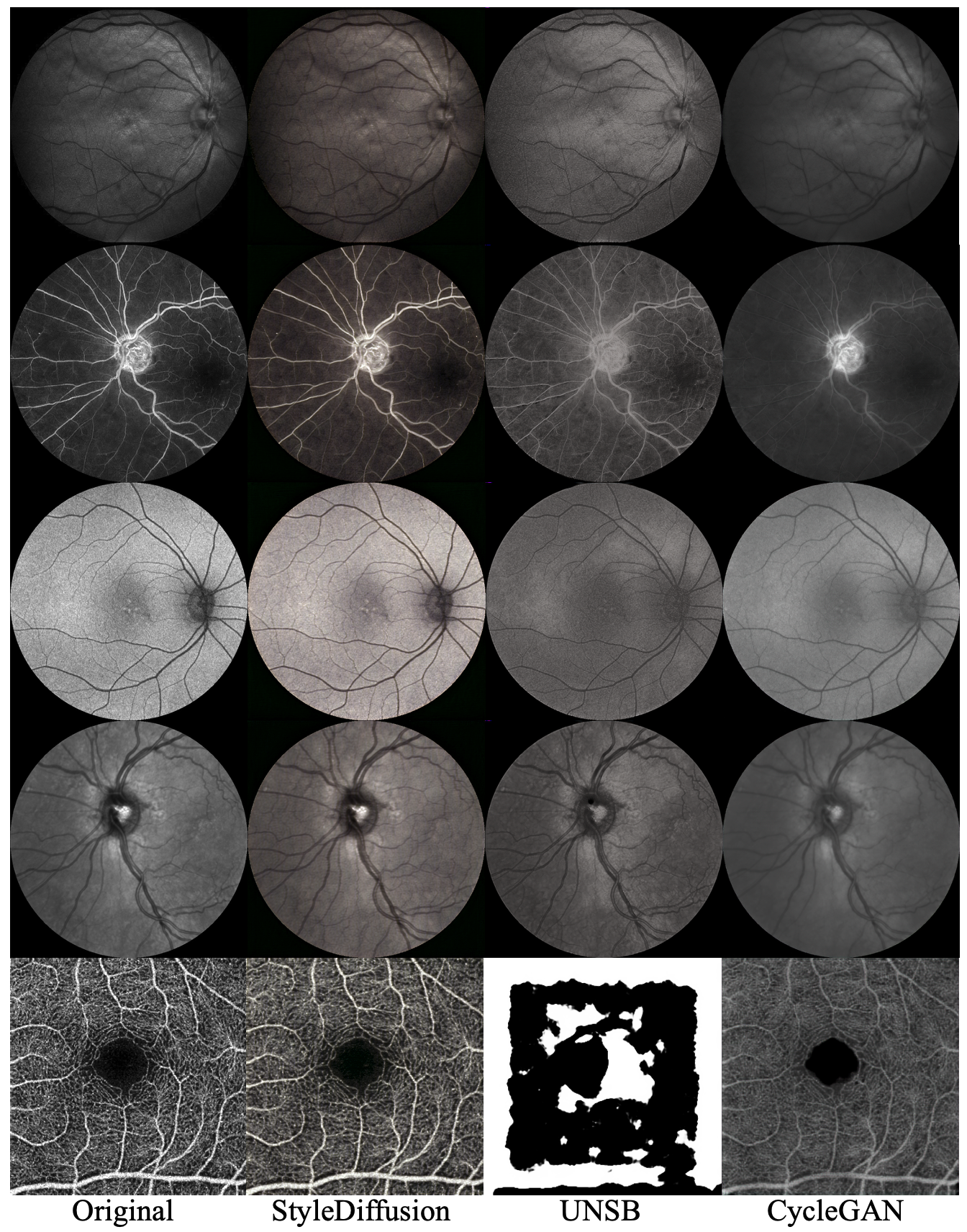}
\vspace{-0.3cm}
\caption{Qualitative examples of the translated images using different image translation methods. From top to bottom: MC, FA, FAF, IR, OCTA. (best viewed by zooming in)}
\label{fig:ab_transfer_examples}
\vspace{-0.6cm}
\end{figure}

\subsection{Comparison with general-purpose models}
\label{ssection:sams}

Although great progress has been made in general-purpose segmentation models in recent years, they struggle with specialized images and structures such as retinal image and curvilinear structures. To further highlight the need for a universal retinal vessel segmentation model, we compare with three recent foundational vision models: SAM2 \cite{SAM2}, Grounded SAM2 \cite{GroundedSAM2}, and the SOTA variant for general medical image segmentation: MedSAM2 \cite{MedSAM2}. For Grounded SAM2, we use a uniform text prompt, `vessels.'. For SAM2 and MedSAM2, we provide manual point prompts with over 100 positive points on the vessels and over 20 negative points on the background for the first image of each dataset. The generated mask is then propagated to all subsequent images. Due to space constraints, we report the results per modality, i.e., average the score over all images (from one or multiple datasets) from each modality. The UWF image is excluded from the MC modality since the extremely high resolution of UWF images makes it infeasible to run these models on our available hardware. The results in \cref{tab:sam_comp} show that our proposed URVSM achieves substantially better performance than these general-purpose segmentation models.

\subsection{Ablation Studies}
Due to the high volume of experimental results, for all the ablation studies, we report the results per modality the same way as described in \cref{ssection:sams}.
\subsubsection{Pipeline}
\label{sssec:pipeline}
In this part, we study how the main components in the proposed pipeline improve the segmentation quality. All the experiments in the ablation studies adopt the same training configurations, data augmentations and network structure, unless otherwise stated. We first run a baseline experiment where neither image translation nor topological learning is applied. In other words, we simply train a segmentation model on the DRIVE training set and apply the model to the multi-modality retinal images. As shown by \cref{fig:ab_step}, the performance of the model is very poor. Both image translation and topological learning improve the performance of the model, while topological learning plays a more dominant role. Optimal performance is achieved by using both I2I and topological learning. An interesting observation is that, for the FA modality, the vessel color is inverted and the model tends to predict the boundary of the vessels as foreground rather than the vessel itself, resulting in a `sandwich' pattern. No other domain adaptation method can resolve this issue except for the proposed topological learning (\cref{fig:DA_FA}).

\subsubsection{Image Translation Method}
\label{sssec:ab_image_translation}
We compare CycleGAN with the two most recent diffusion model-based methods (StyleDiffusion \cite{Trans-StyleDiffusion} and UNSB \cite{Trans-UNSB}). The 19 translated DRIVE images are produced by respective translation method to train the downstream segmentation network.

For StyleDiffusion, we randomly pick only one CF image as the style target and reduce the training set size to the same as that in \cite{Trans-StyleDiffusion} because we find that on a larger training set, StyleDiffusion generates severe artifacts despite our efforts to fine-tune the training hyperparameters for our images. In addition, data augmentation cannot be implemented in StyleDiffusion because of its two-diffusion-stage design. In general, the final segmentation performance by StyleDiffusion is the poorest (\cref{fig:ab_transfer_chart}). Furthermore, similar to the observation in \cite{MMVesseg-Zhang}, where style transfer methods are also used for retinal vessel segmentation, the transferred images tend to include vessel-like artifacts (\cref{fig:ab_transfer_examples}) introduced by the perceptual inductive bias, which in the subsequent step will also result in many artifacts in the segmentation.

For UNSB, the training is performed on the same dataset for training CycleGAN. We use the recommended training settings and network structure in \cite{Trans-UNSB} except for tuning the number of the diffusion steps to 6 to adapt to our data. 

As shown in \cref{fig:ab_transfer_chart}, when comparing the final segmentation performance, CycleGAN demonstrates the best performance for most datasets. It is also observed that the CycleGAN translation results are visually more similar to the target CF images where the sharpness and contrast of the images are weaker than in the other modalities (\cref{fig:ab_transfer_examples}). Furthermore, for FAF where the original images are very noisy, CycleGAN also performs better in reducing the noise and generating a smoother translated image. Additionally, UNSB fails on the OCTA modality while StyleDiffusion and CycleGAN show better generalization ability.

\begin{figure}[]
\centering
\includegraphics[width=0.47\textwidth]{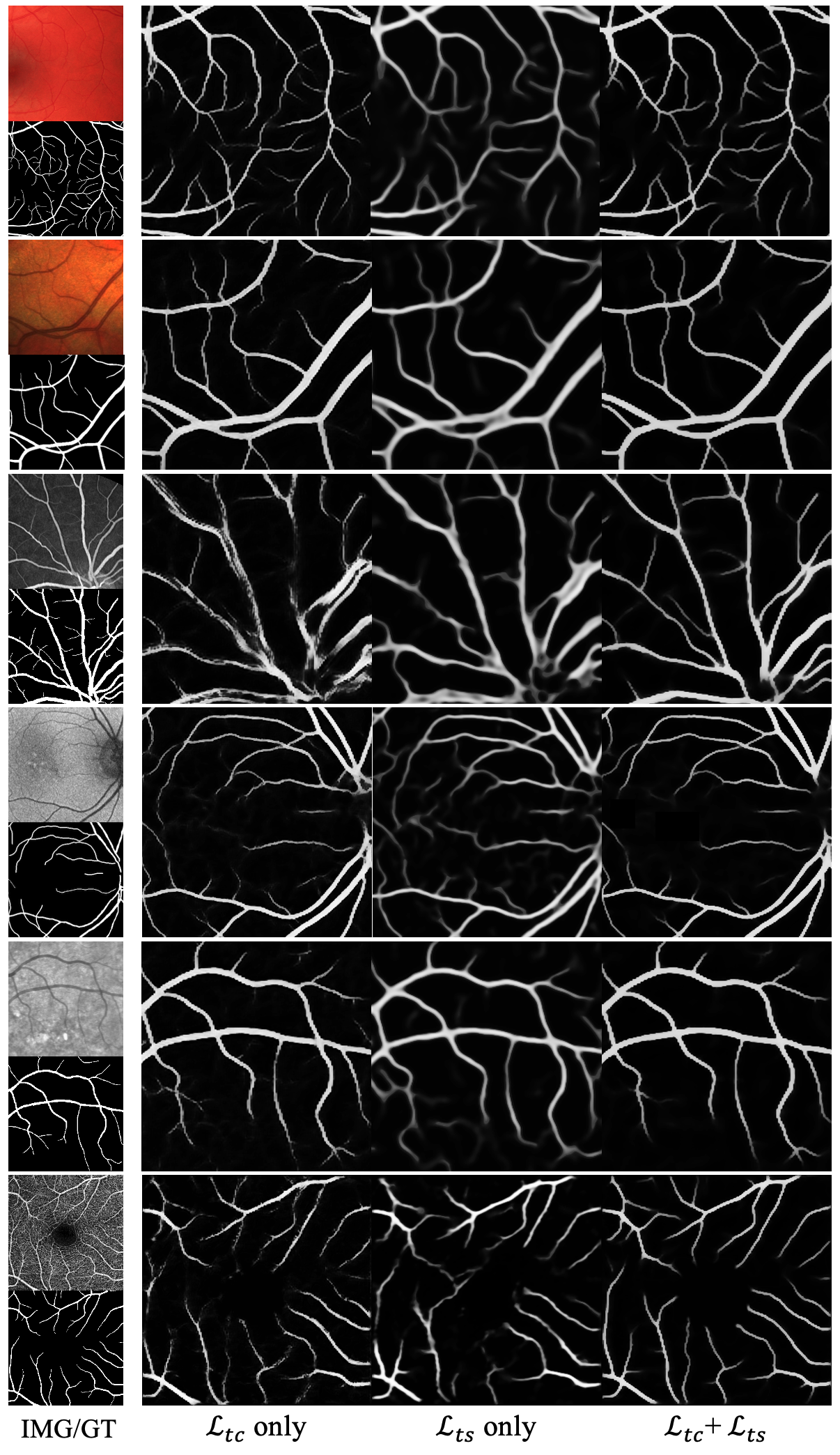}
\vspace{-0.3cm}
\caption{Qualitative results for \cref{sssec:ab_topo}, From top to bottom: CF, MC, FA, FAF, IR, OCTA}
\label{fig:ab_topo}
\vspace{-0.6cm}
\end{figure}

\begin{table}[ht]
  \caption{Comparison between different choices of topological loss functions. While the primary goal is to improve domain generalization, we still choose to report the topological metrics here since they are more relevant to the structural properties of the vessels.}
  \vspace{-0.1cm}
  \footnotesize
  \label{tab:topo_loss}
  \centering
  \scriptsize

  \begin{tabular}{@{}c|cc|>{\columncolor{white!50}}c>{\columncolor{white!50}}c>{\columncolor{white!50}}c>{\columncolor{white!20}}c>{\columncolor{white!20}}c}
    \toprule
    \rowcolor{white} Modality & $\mathcal{L}_{tc}$ & $\mathcal{L}_{ts}$ & Acc. & Dice & clDice & \(\beta_{0}\) $\downarrow$ & \(\beta_{1}\) $\downarrow$ \\


    \hline
    \multirow{4}{*}{\(CF\)}
      & - & - & 96.72 & 73.57 & 74.64 & 114.7 & 35.5 \\
      & $\checkmark$ & - & 96.75 & 73.43 & 74.31 & 149.8 & 38.5 \\
      & - & $\checkmark$ & 96.37 & 71.00 & 72.02 & \textbf{12.1} & \textbf{27.3} \\
      & $\checkmark$ & $\checkmark$ & \textbf{96.80} & \textbf{74.38} & \textbf{75.75} & 61.9 & 29.0 \\

    \hline
    \multirow{4}{*}{\(MC\)}
      & - & - & 96.48 & 77.51 & 84.61 & 87.6 & 8.0 \\
      & $\checkmark$ & - & 96.56 & 77.49 & 83.56 & 160.7 & 6.3 \\
      & - & $\checkmark$ & 96.28 & 76.15 & 83.01 & \textbf{15.6} & 6.6 \\
      & $\checkmark$ & $\checkmark$ & \textbf{96.70} & \textbf{78.38} & \textbf{85.08} & 69.1 & \textbf{6.3} \\

    \hline
    \multirow{4}{*}{\(FA\)}
      & - & - & 90.85 & 40.15 & 36.35 & 233.4 & 17.1 \\
      & $\checkmark$ & - & 94.00 & 58.73 & 60.22 & 325.4 & 98.8 \\
      & - & $\checkmark$ & 95.41 & 69.78 & 72.91 & \textbf{15.6} & 16.0 \\
      & $\checkmark$ & $\checkmark$ & \textbf{96.70} & \textbf{75.00} & \textbf{76.37} & 37.7 & \textbf{15.2} \\

    \hline
    \multirow{4}{*}{\(FAF\)}
      & - & - & 95.90 & 64.45 & 69.03 & 203.9 & 8.9 \\
      & $\checkmark$ & - & 97.03 & 73.46 & 77.88 & 149.7 & 8.8 \\
      & - & $\checkmark$ & 96.52 & 71.00 & 79.07 & \textbf{17.1} & 5.6 \\
      & $\checkmark$ & $\checkmark$ & \textbf{97.30} & \textbf{75.49} & \textbf{81.89} & 21.6 & \textbf{4.5} \\

    \hline
    \multirow{4}{*}{\(IR\)}
      & - & - & 95.36 & 69.77 & 71.34 & 144.0 & 6.7 \\
      & $\checkmark$ & - & 96.01 & 73.48 & 74.66 & 149.2 & 14.6 \\
      & - & $\checkmark$ & 95.69 & 72.09 & 74.69 & \textbf{21.0} & \textbf{5.4} \\
      & $\checkmark$ & $\checkmark$ & \textbf{96.10} & \textbf{74.66} & \textbf{76.30} & 28.8 & 5.6 \\

    \hline
    \multirow{4}{*}{\(OCTA\)}
      & - & - & 89.50 & 33.09 & 33.51 & 83.2 & 7.9 \\
      & $\checkmark$ & - & 94.29 & 62.10 & 64.49 & 102.0 & \textbf{6.6} \\
      & - & $\checkmark$ & 92.22 & 50.89 & 52.45 & 16.8 & 8.3 \\
      & $\checkmark$ & $\checkmark$ & \textbf{95.52} & \textbf{72.54} & \textbf{76.36} & \textbf{13.1} & 7.6 \\
      

  \bottomrule
  \end{tabular}
  \vspace{-0.1cm}
\end{table}

\subsubsection{Topological Segmentation Loss Function}
\label{sssec:ab_topo}
In this part, we compare how learning different combinations of topological features impacts the performance of the URVSM. \cref{tab:topo_loss} and \cref{fig:ab_topo} show that both $\mathcal{L}_{tc}$ and $\mathcal{L}_{ts}$ help overcome the domain gap of non-CF modalities. It is observed that $\mathcal{L}_{tc}$ is unable to address the inverted vessel color in FA images and instead tends to predict the boundary of the vessel, which results in a `sandwich' pattern. In addition, using only $\mathcal{L}_{tc}$ adds saw artifacts and discontinuities to the segmentation. $\mathcal{L}_{ts}$ is able to address the problem of FA modality and reduces the discontinuities along the vessel, however, it is observed that if only $\mathcal{L}_{ts}$ is used, the vessel boundary will be too smoothed and the morphological quality of the segmentation will be poor. With a combination of $\mathcal{L}_{tc}$ and $\mathcal{L}_{ts}$, the saw and smoothing artifacts counteract, and the challenge in predicting FA vessels is effectively mitigated, producing optimal results.

\begin{table}
  \caption{Comparison of using different segmentation network structures for URVSM. Metric reported: Dice score.}
  \vspace{-0.1cm}
  \label{tab:backbone_comp}
  \centering
  \scriptsize

  \begin{tabular}{@{\hspace{2pt}}c@{\hspace{2pt}}|cccccc@{}}
  \toprule
     Backbone & CF & MC & FA & FAF & IR & OCTA \\
  \midrule
  \multicolumn{1}{c|}{U-Net} & 74.38 & \textbf{78.38} & 75.00 & 75.49 & 74.66 & 72.54 \\
  \multicolumn{1}{c|}{FRU-Net} & 75.02 & 77.33 & 73.99 & 76.32 & \textbf{76.00} & 62.85 \\
  \multicolumn{1}{c|}{MCG\&BSA-Net} & 75.53 & 76.94 & 76.23 & 76.34 & 74.18 & 65.99 \\
  \multicolumn{1}{c|}{RCARU-Net} & \textbf{76.02} & 78.00 & 75.70 & 76.07 & 75.28 & 60.27 \\
  \multicolumn{1}{c|}{Wave-Net} & 75.20 & 76.73 & 76.43 & \textbf{76.38} & 75.48 & 71.07 \\
  \multicolumn{1}{c|}{ResDoU-Net} & 76.01 & 77.89 & \textbf{79.13} & 76.15 & 75.49 & \textbf{72.71} \\

  \bottomrule
  \end{tabular}
  \vspace{-0.2cm}
\end{table}

\subsubsection{Segmentation Network Structure}
\label{sssec:ab_backbone}

We also tested and compared several recent network structures in retinal vessel segmentation as the segmentation model backbone for our URVSM, namely, FRUNet \cite{FRR-TSNet}, MCG\&BSA-Net \cite{MLAL-PDSF}, RCARU-Net \cite{Backbone_RCARUNet}, Wave-Net \cite{Backbone_WaveNet} and ResDoU-Net \cite{Backbone_ResdoU-Net}. \cref{tab:backbone_comp} shows that with different backbones, universal retinal vessel segmentation is still achieved, which shows the effectiveness and flexibility of our proposed learning methods. In addition, with some more recent backbones, the overall performance of the URVSM is further improved.

\subsection{Pilot Study on Topological Domain Adaptation}
\begin{figure}[ht]
\centering
\includegraphics[width=0.48\textwidth]{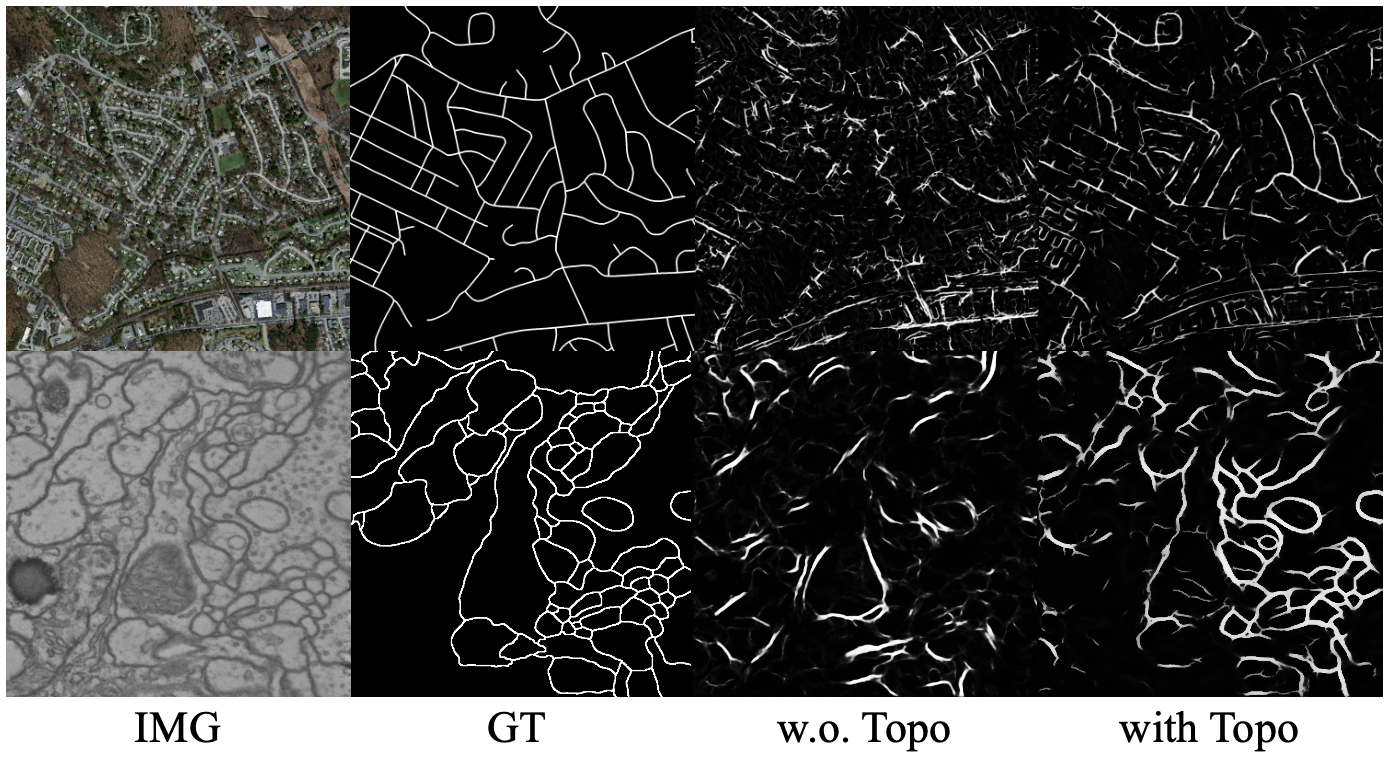}
\vspace{-0.3cm}
\caption{Qualitative examples for the pilot study.}
\label{fig:pilot_study}
\vspace{-0.6cm}
\end{figure}

To further demonstrate the potential of topological learning as a domain adaptation strategy for more general curvilinear structure segmentation, we conducted a preliminary experiment beyond retinal vessel segmentation. Specifically, we directly applied the retinal vessel segmentation model trained with the proposed topological learning (corresponding to “topo” in \cref{fig:ab_step}) and compared it with a baseline model trained without topological learning (“vanilla” in \cref{fig:ab_step}). As reported in \cref{tab:pilot_study} and \cref{fig:pilot_study}, the segmentation model trained with the joint topological learning demonstrates stronger generalization capability than its baseline counterpart.

\begin{table}
  \caption{Performance of retinal vessel segmentation models with/w.o. topological learning on more general curvilinear structures}
  \vspace{-0.1cm}
  \label{tab:pilot_study}
  \centering
  \scriptsize

  \begin{tabular}{@{\hspace{2pt}}c@{\hspace{2pt}}|ccc|ccc@{}}
  \toprule
     Dataset & \multicolumn{3}{c|}{Roads} & \multicolumn{3}{c}{CREMI} \\
         & Acc.$\uparrow$ & Dice$\uparrow$ & clDice$\uparrow$ & Acc.$\uparrow$ & Dice$\uparrow$ & clDice$\uparrow$ \\
  \midrule
  \multicolumn{1}{c|}{w.o. Topo} & 91.02 & 12.81 & 12.05 & 86.68 & 7.55 & 7.41 \\
  \multicolumn{1}{c|}{with Topo} & 92.44 & 24.20 & 23.19 & 87.06 & 12.02 & 11.89 \\

  \bottomrule
  \end{tabular}
  \vspace{-0.2cm}
\end{table}

\section{Conclusions}
\label{sec:Conclusions}
\vspace{-0.2cm}

\subsection{Conclusion}
\label{ssec:conclusion}
In this paper, we propose the universal retinal vessel segmentation task and the very first method for this task, offering a generalizable and more practical solution to the retinal vessel segmentation problem. Compared with the previous literature on retinal vessel segmentation where (1) in most cases only Color Fundus modality is studied and (2) separate models need to be fine-tuned for different modalities and cameras to achieve good performance, our method allows modality/camera-agnostic vessel segmentation for multi-modal retinal images and achieves comparable performance with SOTA fine-tuned methods.

\subsection{Limitations and Future Work}
\label{ssec:limitations and future work}
As demonstrated and discussed in the experimental section, the current version of our URVSM has several limitations that suggest directions for future improvement.

Firstly, there remains a trade-off between universality and performance. While URVSM performs well on primary vessels, it tends to over-prune finer branches, leading to missed segmentation of thin peripheral vessels or capillaries near the macula. This contributes to a performance gap between URVSM and modality-specific SOTA models. The limitation arises from both the detail loss inherent to the image translation process and the regularizing effect of topological losses.

This issue is especially evident in modalities like FA and OCTA, where capillary-level vessels could be visible. As a result, the current model is not suitable for capillary-level analysis, such as FAZ mapping or OCTA perfusion quantification - which we consider as an important direction for future extensions.

Finally, although the model generalizes well to UWF images overall, its robustness to outlier patterns in the peripheral field of UWF scans can still be improved.

\bibliographystyle{IEEEtran}
\bibliography{ref}

\begin{thebibliography}{100}
\providecommand{\url}[1]{#1}
\csname url@samestyle\endcsname
\providecommand{\newblock}{\relax}
\providecommand{\bibinfo}[2]{#2}
\providecommand{\BIBentrySTDinterwordspacing}{\spaceskip=0pt\relax}
\providecommand{\BIBentryALTinterwordstretchfactor}{4}
\providecommand{\BIBentryALTinterwordspacing}{\spaceskip=\fontdimen2\font plus
\BIBentryALTinterwordstretchfactor\fontdimen3\font minus \fontdimen4\font\relax}
\providecommand{\BIBforeignlanguage}[2]{{%
\expandafter\ifx\csname l@#1\endcsname\relax
\typeout{** WARNING: IEEEtran.bst: No hyphenation pattern has been}%
\typeout{** loaded for the language `#1'. Using the pattern for}%
\typeout{** the default language instead.}%
\else
\language=\csname l@#1\endcsname
\fi
#2}}
\providecommand{\BIBdecl}{\relax}
\BIBdecl

\bibitem{Registration-Wang-TIP}
Y.~Wang, J.Zhang, M.~Cavichini, D.~Bartsch, W.~Freeman, T.~Nguyen, and C.~An, ``Robust content-adaptive global registration for multimodal retinal images using weakly supervised deep-learning framework,'' \emph{IEEE Transactions on Image Processing}, vol.~30, pp. 3167--3178, 2021.

\bibitem{Registration-Zhang-TIP}
J.~Zhang, Y.~Wang, J.~Dai, M.~Cavichini, D.-U.G.Bartsch, W.~R. Freeman, T.~Q. Nguyen, and C.~An, ``Two-step registration on multi-modal retinal images via deep neural networks,'' \emph{IEEE Transactions on Image Processing}, vol.~31, pp. 823--838, 2022.

\bibitem{Biometric-Alex}
P.~Alexandru, J.~Kotzerke, and A.~Mertins, ``Robust retina-based person authentication using the sparse classifier,'' in \emph{20th European Signal Processing Conference (EUSIPCO)}, 2012, pp. 1514--1518.

\bibitem{AVCls-Estrada}
R.~Estrada, Rolando, M.~Allingham, and P.~Mettu, ``Retinal artery-vein classification via topology estimation,'' \emph{IEEE Transactions on Medical Imaging}, vol.~34, no.~12, pp. 2518--2534, 2015.

\bibitem{TreeTopology-Estrada}
R.~Estrada, C.~Tomasi, and S.~Schmidler, ``Tree topology estimation,'' \emph{IEEE Transactions on Pattern Analysis and Machine Intelligence}, vol.~37, no.~8, pp. 1688--1701, 2018.

\bibitem{Clinical-Ikram}
M.~Ikram, F.~de~Jong, M.~Bos, J.~Vingerling, A.~Hofman, P.~Koudstaal, P.~de~Jong, and M.~Breteler, ``Retinal vessel diameters and risk of stroke,'' \emph{Neurology}, vol.~66, no.~9, pp. 1339--1343, 2006.

\bibitem{Clinical-Kawasaki}
R.~Kawasaki, J.~Xie, N.~Cheung, E.~Lamoureux, R.~Klein, B.~Klein, M.~Cotch, A.~Sharrett, S.~Shea, and T.~Wong, ``Retinal microvascular signs and risk of stroke: The multi-ethnic study of atherosclerosis,'' \emph{Stroke}, vol.~43, no.~12, pp. 3245--3251, 2012.

\bibitem{Clinical-McGee}
K.~McGeechan, G.~Liew, P.~Macaskill, L.~Irwig, R.~Klein, B.~Klein, J.~Wang, P.~Mitchell, J.~Vingerling, P.~de~Jong, J.~Witteman, M.~Breteler, J.~Shaw, P.~Zimmet, and T.~Wong, ``Prediction of incident stroke events based on retinal vessel caliber: a systematic review and individual-participant meta-analysis,'' \emph{Am J Epidemiol}, vol. 170, no.~11, pp. 1323--1332, 2009.

\bibitem{Dataset-Vampire}
A.~Perez-Rovira, K.~Zutis, J.~P. Hubschman, and E.~Trucco, ``Improving vessel segmentation in ultra-wide field-of-view retinal fluorescein angiograms,'' in \emph{2011 Annual International Conference of the IEEE Engineering in Medicine and Biology Society}, 2011, pp. 2614--2617.

\bibitem{Dataset_DRIVE}
J.~Staal, M.~Abràmoff, M.~N. abd Max~Viergever, and B.~Ginneken, ``Ridge-based vessel segmentation in color images of the retina,'' \emph{IEEE Transactions on Medical Imaging}, vol.~23, no.~4, pp. 501--509, 2004.

\bibitem{MMVesseg-Rodri}
E.~O. Rodrigues, A.~Conci, and P.~Liatsis, ``Element: Multi-modal retinal vessel segmentation based on a coupled region growing and machine learning approach,'' \emph{IEEE Journal of Biomedical and Health Informatics}, vol.~24, no.~12, pp. 3507--3519, 2020.

\bibitem{MMVesseg-Zhang}
J.~Zhang, C.~An, J.~Dai, M.~Amador, D.~Bartsch, S.~Borooah, W.~Freeman, and T.~Nguyen, ``Joint vessel segmentation and deformable registration on multi-modal retinal images based on style transfer,'' in \emph{2019 IEEE International Conference on Image Processing (ICIP)}, 2019, pp. 839--843.

\bibitem{MMVesseg-StyleTransferPeng}
L.~Peng, L.~Lin, P.~Cheng, Z.~Huang, and X.~Tang, ``Unsupervised domain adaptation for cross-modality retinal vessel segmentation via disentangling representation style transfer and collaborative consistency learning,'' in \emph{2022 IEEE 19th International Symposium on Biomedical Imaging (ISBI)}, 2022, pp. 1--5.

\bibitem{SAM2}
\BIBentryALTinterwordspacing
N.~Ravi, V.~Gabeur, Y.-T. Hu, R.~Hu, C.~Ryali, T.~Ma, H.~Khedr, R.~R{\"a}dle, C.~Rolland, L.~Gustafson, E.~Mintun, J.~Pan, K.~V. Alwala, N.~Carion, C.-Y. Wu, R.~Girshick, P.~Doll{\'a}r, and C.~Feichtenhofer, ``Sam 2: Segment anything in images and videos,'' \emph{arXiv preprint arXiv:2408.00714}, 2024. [Online]. Available: \url{https://arxiv.org/abs/2408.00714}
\BIBentrySTDinterwordspacing

\bibitem{GroundedSAM2}
\BIBentryALTinterwordspacing
T.~Ren, S.~Liu, A.~Zeng, J.~Lin, K.~Li, H.~Cao, J.~Chen, X.~Huang, Y.~Chen, F.~Yan, Z.~Zeng, H.~Zhang, F.~Li, J.~Yang, H.~Li, Q.~Jiang, and L.~Zhang, ``Grounded sam: Assembling open-world models for diverse visual tasks,'' 2024. [Online]. Available: \url{https://arxiv.org/abs/2401.14159}
\BIBentrySTDinterwordspacing

\bibitem{MedSAM2}
\BIBentryALTinterwordspacing
J.~Zhu, A.~Hamdi, Y.~Qi, Y.~Jin, and J.~Wu, ``Medical sam 2: Segment medical images as video via segment anything model 2,'' \emph{arXiv preprint arXiv:2408.00874}, 2024. [Online]. Available: \url{https://arxiv.org/abs/2408.00874}
\BIBentrySTDinterwordspacing

\bibitem{Vesseg-MF-Chaudhuri}
S.~Chaudhuri, S.~Chatterjee, N.~Katz, M.~Nelson, and M.~Goldbaum, ``Detection of blood vessels in retinal images using two-dimensional matched filters,'' \emph{IEEE Transactions on Medical Imaging}, vol.~8, no.~3, pp. 263--269, 1989.

\bibitem{Vesseg-MF-Kovacs}
G.~Kovács and A.~Hajdu, ``A self-calibrating approach for the segmentation of retinal vessels by template matching and contour reconstruction.'' \emph{Medical Image Analysis}, vol.~29, pp. 24--46, 2016.

\bibitem{Dataset-HRF}
J.~Odstrcilik, R.~Kolar, A.~Budai, J.~Hornegger, J.~Jan, G.~J, T.~Kubena, P.~Cernosek, O.~Svoboda, and E.~Angelopoulou, ``Retinal vessel segmentation by improved matched filtering: evaluation on a new high-resolution fundus image database.'' \emph{IET Image Processing}, vol.~7, no.~4, pp. 373--383, 2013.

\bibitem{Vesseg-Tracking-Delibasis}
K.~Delibasis, A.~Kechriniotis, C.~Tsonos, and N.~Assimakis, ``Automatic model-based tracing algorithm for vessel segmentation and diameter estimation,'' \emph{Comput. Methods Programs Biomed.}, vol. 100, no. 108-122, 2010.

\bibitem{Vesseg-Tracking-Lin}
K.~Lin, C.~Tsai, C.~Tsai, M.~Sofka, S.~Chen, and W.~Lin, ``Retinal vascular tree reconstruction with anatomical realism.'' \emph{IEEE Transactions on Biomedical Engineering.}, vol.~59, no.~12, pp. 3337--3347, 2012.

\bibitem{Vesseg-Morph-Graz}
M.~Fraz, A.~Basit, and S.~Barman, ``Application of morphological bit planes in retinal blood vessel extraction.'' \emph{J Digit Imaging}, vol.~26, no.~2, pp. 274--286, 2013.

\bibitem{Vesseg-Morph-Imani}
E.~Imani, M.~Javidi, and H.~Pourreza, ``Improvement of retinal blood vessel detection using morphological component analysis.'' \emph{Comput. Methods Programs Biomed}, vol. 118, no.~3, pp. 263--279, 2015.

\bibitem{Vesseg-SVM-Tang}
Z.~Tang, J.~Zhang, and W.~Gui, ``Selective search and intensity context based retina vessel image segmentation.'' \emph{J. Med. Syst.}, vol.~41, no.~3, p.~47, 2017.

\bibitem{Vesseg-RF-Wang}
S.~Wang, Y.~Yin, G.~Cao, B.~Wei, Y.~Zheng, and G.~Yang, ``Hierarchical retinal blood vessel segmentation based on feature and ensemble learning.'' \emph{Neurocomputing}, vol. 149, pp. 708--717, 2015.

\bibitem{Vesseg-Adaboost-Memari}
M.~N, A.~Ramli, M.~Saripan, S.~Mashohor, and M.~Moghbel, ``A discriminatively trained fully connected conditional random field model for blood vessel segmentation in fundus images.'' \emph{IEEE Transactions on Biomedical Engineering}, vol.~64, no.~1, pp. 16--27, 2017.

\bibitem{Vesseg-GMM-Roy}
S.~Roychowdhury, D.~D. Koozekanani, and K.~K. Parhi, ``Blood vessel segmentation of fundus images by major vessel extraction and subimage classification,'' \emph{IEEE Journal of Biomedical and Health Informatics}, vol.~19, no.~3, pp. 1118--1128, 2015.

\bibitem{Vesseg-FCM-Neto}
L.~Neto, G.~Ramalho, J.~Neto, R.~Veras, and F.~Medeiros, ``Blood vessel segmentation of fundus images by major vessel extraction and subimage classification,'' \emph{An unsupervised coarse-to-fine algorithm for blood vessel segmentation in fundus images.}, vol.~78, p. 182–192, 2017.

\bibitem{Vesseg-CNN-DRIU}
L.~Maninis, J.~Pont-Tuset, P.~Arbeláez, and L.~V. Gool, ``Deep retinal image understanding,'' in \emph{2016 Medical Image Computing and Computer-Assisted Intervention (MICCAI)}, 2016, pp. 140--148.

\bibitem{Others-VGG}
A.~Z. Karen~Simonyan, ``Very deep convolutional networks for large-scale image recognition,'' 2014, arXiv preprint arXiv:1409.1556.

\bibitem{Others-U-Net}
O.~Ronneberger, P.~Fischer, and T.~Brox, ``U-net: Convolutional networks for biomedical image segmentation,'' in \emph{Medical Image Computing and Computer-Assisted Intervention}, 2015, pp. 234--241.

\bibitem{Vesseg-CNN-CENet}
Z.~Gu, J.~Cheng, H.~Fu, K.~Zhou, H.~Hao, Y.~Zhao, T.~Zhang, S.~Gao, and J.~Liu, ``Ce-net: Context encoder network for 2d medical image segmentation,'' \emph{IEEE Transactions on Medical Imaging}, vol.~38, no.~10, pp. 2281--2292, 2019.

\bibitem{Vesseg-CNN-CS2Net}
L.~Mou, Y.~Zhao, H.~Fu, Y.~Liu, J.~Cheng, Y.~Zheng, P.~Su, J.~Yang, L.~Chen, A.~F. Frangi, M.~Akiba, and J.~Liu, ``Cs2-net: Deep learning segmentation of curvilinear structures in medical imaging,'' \emph{Medical Image Analysis}, vol.~67, p. 101874, 2019.

\bibitem{Vesseg-CNN-GlobalTransformer}
Y.~Li, Y.~Zhang, J.-Y. Liu, K.~Wang, K.~Zhang, G.-S. Zhang, X.-F. Liao, and G.~Yang, ``Global transformer and dual local attention network via deep-shallow hierarchical feature fusion for retinal vessel segmentation,'' \emph{IEEE Transactions on Cybernetics}, vol.~53, no.~9, pp. 5826--5839, 2023.

\bibitem{Vesseg-CNN-MCDAUNet}
W.~Zhou, W.~Bai, J.~Ji, Y.~Yi, N.~Zhang, and W.~Cui, ``Dual-path multi-scale context dense aggregation network for retinal vessel segmentation,'' \emph{Computers in Biology and Medicine}, vol. 164, p. 107269, 2019.

\bibitem{Vesseg-CNN-MCGNet}
J.~Liu, Z.~J, J.~Xiao, G.~Zhao, P.~Xu, Y.~Yang, and S.~Gong, ``Unsupervised domain adaptation multi-level adversarial learning-based crossing-domain retinal vessel segmentation.'' \emph{Computers in Biology and Medicine}, vol. 178, p. 108759, 2024.

\bibitem{Vesseg-CNN-RCARUNet}
W.~Ding, Y.~Sun, J.~Huang, H.~Ju, C.~Zhang, G.~Yang, and C.~Lin, ``Rcar-unet: Retinal vessel segmentation network algorithm via novel rough attention mechanism,'' \emph{Information Sciences}, vol. 657, p. 120007, 2024.

\bibitem{Vesseg-CNN-WaveNet}
Y.~Liu, J.~Shen, L.~Yang, H.~Yu, and G.~Bian, ``Wave-net: A lightweight deep network for retinal vessel segmentation from fundus images,'' \emph{Computers in Biology and Medicine}, vol. 152, p. 106341, 2023.

\bibitem{Vesseg-CNN-DUNet}
Q.~Jin, Z.~Meng, T.~D. Pham, Q.~Chen, L.~Wei, and R.~Su, ``Dunet: A deformable network for retinal vessel segmentation,'' \emph{Knowledge-Based Systems}, vol. 178, pp. 149--162, 2019.

\bibitem{Topo-TCLoss}
Y.~Qi, Y.~He, X.~Qi, Y.~Zhang, and G.~Yang, ``Dynamic snake convolution based on topological geometric constraints for tubular structure segmentation,'' in \emph{IEEE/CVF International Conference on Computer Vision}, 2023, pp. 6070--6079.

\bibitem{Vesseg-CNN-WSDMF}
Y.~Tan, K.~Yang, S.~Zhao, W.~J, L.~Liu, and Y.~Li, ``Deep matched filtering for retinal vessel segmentation,'' \emph{Knowledge-Based Systems}, vol. 283, p. 111185, 2024.

\bibitem{Vesseg-CNN-WNet}
A.~Galdran, A.~Anjos, J.~Dolz, H.~Chakor, H.~Lombaert, and I.~Ayed, ``State-of-the-art retinal vessel segmentation with minimalistic models.'' \emph{Scientific Reports}, vol.~12, no.~1, p. 6174, 2022.

\bibitem{Vesseg-CNN-FANet}
N.~Tomar, D.~Jha, M.~Riegler, H.~Johansen, D.~Johansen, J.~Rittscher, P.~Halvorsen, and S.~Ali, ``Fanet: A feedback attention network for improved biomedical image segmentation.'' \emph{IEEE Transactions on Neural Networks and Learning Systems}, vol.~34, no.~11, pp. 9375--9388, 2022.

\bibitem{Trans-CycleGAN}
J.-Y. Zhu, T.~Park, P.~Isola, and A.~A. Efros, ``Unpaired image-to-image translation using cycle-consistent adversarial networks,'' in \emph{IEEE/CVF International Conference on Computer Vision}, 2017, pp. 2223--2232.

\bibitem{Trans-CcGAN}
H.~Fu, M.~Gong, C.~Wang, K.~Batmanghelich, K.~Zhang, and D.~Tao, ``Geometry- consistent generative adversarial networks for one-sided unsupervised domain mapping.'' in \emph{IEEE/CVF Conference on Computer Vision and Pattern Recognition}, 2019, pp. 2427--2436.

\bibitem{Trans-Mutual}
S.~Benaim and L.~W.~O. sided unsupervised~domain mapping, ``One-sided unsupervised domain mapping.'' in \emph{Advances in Neural Information Processing Systems (NIPS)}, 2019.

\bibitem{Trans-CUT}
T.~Park, A.~Efros, R.~Zhang, and J.-Y. Zhu, ``Contrastive learning for unpaired image-to-image translation.'' in \emph{European Conference on Computer Vision (ECCV)}, 2020, pp. 319--345.

\bibitem{Trans-MUIST}
X.~Huang, M.~Liu, S.~Belongie, and J.~Kautz, ``Multimodal unsupervised image-to-image translation.'' in \emph{European Conference on Computer Vision (ECCV)}, 2018, pp. 172--189.

\bibitem{Trans-StarGAN}
Y.~YChoi, M.~Choi, M.~Kim, J.-W. Ha, S.~Kim, and J.~Choo, ``Stargan: Unified generative adversarial networks for multi-domain image-to-image translation,'' in \emph{Proceedings of the IEEE Conference on Computer Vision and Pattern Recognition (CVPR)}, June 2018.

\bibitem{Others-Diffusion}
J.~Sohl-Dickstein, E.~Weiss, N.~Maheswaranathan, and S.~Ganguli, ``Deep unsupervised learning using nonequilibrium thermodynamics,'' in \emph{International Conference on Machine Learning, PMLR}, 2015, pp. 2256--2265.

\bibitem{Others-DDPM}
J.~Ho, A.~Jain, and P.~Abbeel, ``Denoising diffusion probabilistic models,'' in \emph{Advances in Neural Information Processing Systems}, 2020, pp. 6840--6851.

\bibitem{Others-DDIM}
\BIBentryALTinterwordspacing
J.~Song, C.~Meng, and S.~Ermon, ``Denoising diffusion implicit models,'' \emph{arXiv:2010.02502}, October 2020. [Online]. Available: \url{https://arxiv.org/abs/2010.02502}
\BIBentrySTDinterwordspacing

\bibitem{Trans-DDIB}
X.~Su, J.~Song, C.~Meng, and S.~Ermon, ``Dual diffusion implicit bridges for image-to-image translation,'' in \emph{International Conference on Learning Representations}, 2023.

\bibitem{Trans-ENOT}
N.~Gushchin, A.~Kolesov, A.~Korotin, D.~Vetrov, and E.~Burnaev, ``Entropic neural optimal transport via diffusion processes,'' in \emph{Advances in Neural Information Processing Systems}, 2023.

\bibitem{Trans-SynDiff}
M.~Ozbey, O.~Dalmaz, S.~Dar, H.~Bedel, S.~Ozturk, A.~Gungor, and T.~Cukur, ``Unsupervised medical image translation with adversarial diffusion models.'' \emph{IEEE Transactions on Medical Imaging}, vol.~42, no.~12, pp. 3524--3539, 2023.

\bibitem{Trans-UNSB}
B.~Kim, G.~Kwon, K.~Kim, and J.~C. Ye, ``Unpaired image-to-image translation via neural schrödinger bridge,'' in \emph{International Conference on Learning Represenations (ICLR)}, 2024.

\bibitem{Trans-NAAS}
L.~Gatys, ``A neural algorithm of artistic style,'' 2015, arXiv preprint arXiv:1508.06576.

\bibitem{Trans-PerceptualLoss}
J.~Johnson, A.~Alahi, and F., ``Perceptual losses for real-time style transfer and super-resolution,'' in \emph{European Conference on Computer Vision (ECCV)}, 2016, pp. 694--711.

\bibitem{Trans-EFDM}
Y.~Zhang, M.~Li, R.~Li, K.~Jia, and L.~Zhang, ``Exact feature distribution matching for arbitrary style transfer and domain generalization,'' in \emph{Proceedings of the IEEE/CVF Conference on Computer Vision and Pattern Recognition (CVPR)}, 2022, pp. 8035--8045.

\bibitem{Trans-StyTr2}
Y.~Deng, F.~Tang, W.~Dong, C.~Ma, X.~Pan, L.~Wang, and C.~Xu, ``Stytr2: Image style transfer with transformers,'' in \emph{Proceedings of the IEEE/CVF Conference on Computer Vision and Pattern Recognition (CVPR)}, 2022, pp. 11\,326--11\,336.

\bibitem{Trans-StyleDiffusion}
Z.~Wang, L.~Zhao, and W.~Xing, ``Stylediffusion: Controllable disentangled style transfer via diffusion models,'' in \emph{Proceedings of the IEEE/CVF International Conference on Computer Vision (ICCV)}, October 2023, pp. 7677--7689.

\bibitem{Topo-PTLoss}
A.~Mosinska, P.~Marquez-Neila, M.~K. ski, and P.~Fua, ``Beyond the pixel-wise loss for topology-aware delineation,'' in \emph{IEEE/CVF International Conference on Computer Vision and Pattern Recognition}, 2018, pp. 3136--3145.

\bibitem{Others-ImageNet}
J.~Deng, W.~Dong, R.~Socher, L.-J. Li, L.~Kai, and F.-F. Li, ``Imagenet: A large-scale hierarchical image database,'' in \emph{IEEE Conference on Computer Vision and Pattern Recognition}, 2009, pp. 248--255.

\bibitem{Topo-GlandSeg}
H.~Wang, M.~Xian, and A.~Vakanski, ``Ta-net: Topology-aware network for gland segmentation,'' in \emph{IEEE/CVF Winter Conference on Applications of Computer Vision}, 2022, pp. 1556--1564.

\bibitem{Topo-clDice}
S.~Shit, J.~C. Paetzold, A.~Sekuboyina, I.~Ezhov, A.~Unger, A.~Zhylka, J.~P.~W. Pluim, U.~Bauer, and B.~H. Menze, ``cldice - a novel topology-preserving loss function for tubular structure segmentation,'' in \emph{IEEE/CVF International Conference on Computer Vision and Pattern Recognition}, 2021, pp. 16\,560--16\,569.

\bibitem{Topo-WTLoss}
X.~Hu, L.~Fuxin, D.~Samaras, and C.~Chen, ``Topology-preserving deep image segmentation,'' in \emph{Advances in Neural Information Processing Systems}, 2019.

\bibitem{Topo-WTLoss3D}
C.-C. Wong and C.-M. Vong, ``Persistent homology based graph convolution network for fine-grained 3d shape segmentation,'' in \emph{IEEE/CVF International Conference on Computer Vision}, 2021, pp. 7098--7107.

\bibitem{Topo-CMR}
N.~Byrne, J.~R. Clough, I.~Valverde, G.~Montana, and A.~P. King, ``A persistent homology-based topological loss for cnn-based multiclass segmentation of cmr,'' \emph{IEEE Transactions on Medical Imaging}, vol.~42, no.~1, pp. 3--14, 2023.

\bibitem{Topo-TPAMI}
J.~R. Clough, N.~Byrne, I.~Oksuz, V.~A. Zimmer, J.~A. Schnabel, and A.~P. King, ``A topological loss function for deep-learning based image segmentation using persistent homology,'' \emph{IEEE Transactions on Pattern Analysis and Machine Intelligence}, vol.~44, no.~12, pp. 8766--8778, 2022.

\bibitem{Topo-MLLoss}
H.~He, J.~Wang, P.~Wei, F.~Xu, X.~Ji, C.~Liu, and J.~Chen, ``Toposeg: Topology-aware nuclear instance segmentation,'' in \emph{IEEE/CVF International Conference on Computer Vision}, 2023, pp. 21\,307--21\,316.

\bibitem{Topo-BMLoss}
N.~Stucki, J.~Paetzold, S.~Shit, B.~Menze, and U.~Bauer, ``Topologically faithful image segmentation via induced matching of persistence barcodes,'' in \emph{International Conference on Machine Learning, PMLR}, 2023, pp. 32\,698--32\,727.

\bibitem{Topo-SATLoss}
B.~Wen, H.~Zhang, D.~Bartsch, W.~Freeman, T.~Nguyen, and C.~An, ``Topology-preserving image segmentation with spatial-aware persistent feature matching,'' 2024, arXiv preprint arXiv:2412.02076.

\bibitem{Topo-Wassersteinforpersistence}
M.~Kerber, D.~Morozov, and A.~Nigmetov, ``Geometry helps to compare persistence diagrams,'' 2016, arXiv:1606.03357.

\bibitem{Topo-Hausdorff_distance}
F.~Hausdorff, Ed., \emph{Grundzüge der mengenlehre}.\hskip 1em plus 0.5em minus 0.4em\relax Leipzig Viet, 1914.

\bibitem{Topo-Induced_matching}
U.~Bauer and M.~Lesnick, ``Induced matchings of barcodes and the algebraic stability of persistence,'' in \emph{13th Annual symposium on Computational geometry}, 2015, pp. 355--364.

\bibitem{MLAL-PDSF}
J.~Liu, J.~Zhao, J.~Xiao, G.~Zhao, P.~Xu, Y.~Yang, and S.~Gong, ``Unsupervised domain adaptation multi-level adversarial learning-based crossing-domain retinal vessel segmentation.'' \emph{Computers in Biology and Medicine}, vol. 178, p. 108759, 2024.

\bibitem{FRR-TSNet}
K.~Yue1, L.~Zhan2, and Z.~Wang, ``Unsupervised domain adaptation teacher–student network for retinal vessel segmentation via full-resolution refined model,'' \emph{Scientific Reports}, vol.~15, no.~1, p. 2038, 2025.

\bibitem{UA-Vesseg-AMCD}
J.~Zhuang, Z.~Chen, J.~Zhang, D.~Zhang, and Z.~Cai, ``Domain adaptation for retinal vessel segmentation using asymmetrical maximum classifier discrepancy.'' in \emph{ACM turing celebration conference}, 2019, pp. 1--6.

\bibitem{Others-GAN}
I.~Goodfellow, J.~Pouget-Abadie, M.~Mirza, B.~Xu, D.~Warde-Farley, S.~Ozair, A.~Courville, and Y.~Bengio, ``Generative adversarial nets,'' in \emph{Advances in Neural Information Processing Systems}, vol.~27, 2014.

\bibitem{Others-LSGAN}
X.~Mao, Q.~Li, H.~Xie, R.~Y. Lau, Z.~Wang, and S.~Paul~Smolley, ``Least squares generative adversarial networks,'' in \emph{Proceedings of the IEEE International Conference on Computer Vision (ICCV)}, 2017.

\bibitem{Topo-cc}
H.~Wagner, C.~Chen, and E.~Vucini, ``Efficient computation of persistent homology for cubical data,'' pp. 91--106, 2011, in Topological methods in data analysis and visualization II: theory, algorithms, and applications.

\bibitem{Topo-CompTopoBook}
H.~Edelsbrunner and J.~Harer, Eds., \emph{Computational Topology: An Introduction}.\hskip 1em plus 0.5em minus 0.4em\relax American Mathematical Society, 2022.

\bibitem{Dataset-STARE}
A.~Hoover, V.~Kouznetsova, and M.~Goldbaum, ``Locating blood vessels in retinal images by piecewise threshold probing of a matched filter response,'' \emph{IEEE Transactions on Medical Imaging}, vol.~19, no.~3, pp. 203--210, 2000.

\bibitem{Dataset-ChaseDB1}
M.~M. Fraz, P.~Remagnino, A.~Hoppe, B.~Uyyanonvara, A.~R. Rudnicka, C.~G. Owen, and S.~A. Barman, ``An ensemble classification-based approach applied to retinal blood vessel segmentation,'' \emph{IEEE Transactions on Biomedical Engineering}, 2012.

\bibitem{Dataset-IOSTAR}
J.~Zhang, B.~Dashtbozorg, E.~Bekkers, J.~P.~W. Pluim, R.~Duits, and B.~M. ter Haar~Romeny, ``Robust retinal vessel segmentation via locally adaptive derivative frames in orientation scores,'' \emph{IEEE Transactions on Medical Imaging}, vol.~35, no.~12, pp. 2631--2644, 2016.

\bibitem{Dataset-OCTA500}
H.~Ning, C.~Wang, and X.~Chen, ``An accurate and efficient neural network for octa vessel segmentation and a new dataset,'' in \emph{IEEE International Conference on Acoustics, Speech and Signal Processing (ICASSP)}, 2024, pp. 1966--1970.

\bibitem{Dataset-PRIMEFP20}
L.~Ding, A.~E. Kuriyan, R.~S. Ramchandran, C.~C. Wykoff, and G.~Sharma, ``Weakly-supervised vessel detection in ultra-widefield fundus photography via iterative multi-modal registration and learning,'' \emph{IEEE Transactions on Medical Imaging}, vol.~40, no.~10, pp. 2748--2758, 2021.

\bibitem{Dataset-CFFA}
S.~H.~M. Alipour, H.~Rabbani, and M.~R. Akhlaghi, ``Diabetic retinopathy grading by digital curvelet transform,'' \emph{Comput. Math. Methods Med.}, vol. 2012, no.~1, p. 1607–1614, 2016.

\bibitem{data_road}
V.~Mnih, ``Machine learning for aerial image labeling,'' Ph.D. dissertation, University of Toronto (Canada), 2013.

\bibitem{data_cremi}
J.~Funke, F.~Tschopp, W.~Grisaitis, A.~Sheri-dan, C.~Singh, S.~Saalfeld, and S.~C. Turaga, ``Large scale image segmentation with structured loss based deep learning for connectome reconstruction,'' \emph{IEEE Transactions on Pattern Analysis and Machine Intelligence}, vol.~41, no.~7, pp. 1669--1680, 2019.

\bibitem{gudhi:urm}
\BIBentryALTinterwordspacing
T.~G. Project, \emph{GUDHI User and Reference Manual}, 3rd~ed.\hskip 1em plus 0.5em minus 0.4em\relax GUDHI Editorial Board, 2024. [Online]. Available: \url{https://gudhi.inria.fr/doc/3.10.1/}
\BIBentrySTDinterwordspacing

\bibitem{UWF_SerpMamba}
H.~Wang, Y.~Chen, W.~Chen, H.~Xu, H.~Zhao, B.~Sheng, H.~Fu, G.~Yang, and L.~Zhu, ``Serp-mamba: Advancing high-resolution retinal vessel segmentation with selective state-space model,'' \emph{IEEE Transactions on Medical Imaging}, 2025.

\bibitem{UNet++}
Z.~Zhou, M.~M.~R. Siddiquee, N.~Tajbakhsh, and J.~Liang, ``Unet++: A nested u-net architecture for medical image segmentation,'' in \emph{Medical Image Analysis and Multimodal Learning for Clinical Decision Support : 4th International Workshop}, 2018, pp. 3--11.

\bibitem{FARGO}
L.~Peng, L.~Lin, P.~Cheng, Z.~Wang, and X.~Tang, ``Fargo: A joint framework for faz and rv segmentation from octa images,'' in \emph{Medical Image Computing and Computer Assisted Intervention (MICCAI)}, 2021.

\bibitem{UWF_LKMUNet}
J.~Wang, J.~Chen, D.~Z. Chen, and J.~Wu, ``Lkm-unet: Large kernel vision mamba unet for medical image segmentation,'' in \emph{proceedings of Medical Image Computing and Computer Assisted Intervention}, vol. LNCS 15008, 2024.

\bibitem{UWF_Swin-UMamba}
\BIBentryALTinterwordspacing
J.~Liu, H.~Yang, H.-Y. Zhou, Y.~Xi, L.~Yu, Y.~Yu, Y.~Liang, G.~Shi, S.~Zhang, H.~Zheng, and S.~Wang, ``Grounded sam: Assembling open-world models for diverse visual tasks,'' 2024. [Online]. Available: \url{https://arxiv.org/abs/2401.03302}
\BIBentrySTDinterwordspacing

\bibitem{UWF_EMNet}
A.~Chang, J.~Zeng, R.~Huang, and D.~Ni, ``Em-net: Efficient channel and frequency learning with mamba for 3d medical image segmentation,'' in \emph{International Conference on Medical Image Computing and Computer-Assisted Intervention}, 2024, pp. 266--275.

\bibitem{Backbone_RCARUNet}
W.~Dinga, Y.~Sun, J.~Huang, H.~Ju, C.~Zhang, G.~Yang, and C.-T. Lin, ``Rcar-unet: Retinal vessel segmentation network algorithm via novel rough attention mechanism,'' \emph{Information Sciences}, vol. 657, p. 120007, 2024.

\bibitem{Backbone_WaveNet}
Y.~Liu, J.~Shena, L.~Yanga, H.~Yua, and G.~Bian, ``Wave-net: A lightweight deep network for retinal vessel segmentation from fundus images,'' \emph{Computers in Biology and Medicine}, vol. 152, p. 106341, 2023.

\bibitem{Backbone_ResdoU-Net}
Y.~Liua, J.~Shena, L.~Yanga, G.~Biana, and H.~Yu, ``Resdo-unet: A deep residual network for accurate retinal vessel segmentation from fundus images,'' \emph{Biomedical Signal Processing and Control}, vol.~79, p. 104087, 2023.

\end{thebibliography}

\vfill

\end{document}